\documentclass[twocolumn,aps,pra,superscriptaddress,longbibliography]{revtex4-1}
\usepackage{color}
\usepackage{amsmath}
\usepackage{amssymb}
\usepackage{graphicx}
\usepackage{tikz}
\usepackage{multirow}
\usepackage{float}
\usepackage{bbm}
\usepackage{dsfont}
\usepackage{cases}
\usepackage{mathtools}

\usepackage{epsfig}
\usepackage{verbatim}
\usepackage{array}
\usepackage{setspace}

\usepackage{scalerel}
\usepackage{stackengine,wasysym}
\usepackage{empheq, nccmath}

\usepackage[T1]{fontenc}

\usepackage[unicode=true,
bookmarks=true,bookmarksnumbered=false,bookmarksopen=false,
breaklinks=false,pdfborder={0 0 1},backref=false,colorlinks=true]
{hyperref}
\hypersetup{
	linkcolor=magenta, urlcolor=blue, citecolor=blue, pdfstartview={FitH}, hyperfootnotes=false, unicode=true}

\makeatletter
\@ifundefined{textcolor}{}
{%
	\definecolor{BLACK}{gray}{0}
	\definecolor{WHITE}{gray}{1}
	\definecolor{RED}{rgb}{1,0,0}
	\definecolor{GREEN}{rgb}{0,1,0}
	\definecolor{BLUE}{rgb}{0,0,1}
	\definecolor{CYAN}{cmyk}{1,0,0,0}
	\definecolor{MAGENTA}{cmyk}{0,1,0,0}
	\definecolor{YELLOW}{cmyk}{0,0,1,0}
}

\begin{document}

\preprint{APS/123-QED}

\title{Solution of the Ising model with Brascamp-Kunz boundary conditions by the transfer matrix method}

\author{De-Zhang Li}
\address{Quantum Science Center of Guangdong-Hong Kong-Macao Greater Bay Area, Shenzhen 518045, China}

\author{Xin Wang}
\thanks{Corresponding author: x.wang@cityu.edu.hk}
\address {Department of Physics, City University of Hong Kong, Hong Kong SAR, China}
\affiliation{City University of Hong Kong Shenzhen Research Institute, Shenzhen 518057, China}

\begin{abstract}
The square lattice Ising model under the Brascamp-Kunz boundary conditions is a well-known exactly solvable lattice model. The exact solution of this system has been derived within the framework of Pfaffian-type method. In this paper we provide a derivation for the solution by the Schultz-Mattis-Lieb method in the transfer matrix formalism. We set special interactions on the boundaries and take certain limit of these interactions, so that the system under the Brascamp-Kunz boundary conditions is transformed into another system under the toroidal boundary conditions. The Schultz-Mattis-Lieb method is applied to the mapping system and the partition function is exactly solved in the fermionic representation. The Fisher zeros are analytically calculated and the physical critical point is identified. We also discuss the difference between the transfer matrix approaches to the Brascamp-Kunz and to the toroidal boundary conditions. Our work introduces a member to the family of transfer-matrix-based studies for Ising model under various boundary conditions. 
\end{abstract}

\keywords{Ising model, Brascamp-Kunz boundary conditions, Jordan-Wigner transformation, fermionic representation}
\maketitle

\section{Introduction} \label{intro}
Ising model is one of the most fundamental systems in the field of statistical lattice models. First introduced by Lenz and Ising in the 1920s \cite{RN75, RN441, RN312}, it has a long history of more than one hundred years. In the early studies the partition function of one-dimensional Ising model was exactly solved \cite{RN75}, and theoretical approximations for higher dimensions were proposed \cite{RN460, RN236, RN237}. The first exact solution in two dimensions is Onsager's solution for the square lattice in the absence of a magnetic field \cite{RN72}. Since then, the zero-field models on some other two-dimensional lattices, including the honeycomb \cite{RN122, RN123}, the triangular \cite{RN81}, the Kagom\'e \cite{RN121, RN82}, and the checkerboard lattices \cite{RN58}, have been exactly solved. Lee and Yang first found the exact solution for the square lattice Ising model in an imaginary field $i(\pi/2)k_B T$ \cite{RN57}, in their work on Lee-Yang zeros. The zero-field and imaginary-field cases correspond to the two points of intersection of the unit circle and the real axis in the complex $x=e^{ -2\beta H_{\rm{ex}} }$ plane, where $H_{\rm{ex}}$ represents the field. The imaginary field case is unique as the solution of Ising model in a non-zero field is generally an unsolved problem. For the models on other two-dimensional lattices mentioned above, the imaginary field case has also been solved \cite{RN67, RN68, RN51, RN274, RN558}. 

In this paper we focus on the square lattice Ising model in a zero field. Onsager derived the exact solution of this model in the thermodynamic limit by the transfer matrix method \cite{RN72}. This achievement is a milestone in the research of statistical physics of lattice systems, in particular the exactly solvable models \cite{RN49, RN465}. Kaufman simplified Onsager's method using the spinor representation theory, to give the finite lattice partition function under the toroidal (periodic in both directions) boundary conditions (BCs) \cite{RN73}. In the framework of transfer matrix method, in addition to Kaufman's approach, there are also various notable simplications \cite{RN267, RN76, RN668, RN666}. After Kaufman's solution was presented, a quite different type of approach, now known as the combinatorial formulation, was proposed and developed \cite{RN74, RN272, RN611, RN662, RN663, RN672, RN664, RN673}. One important type of variant of the combinatorial formulation is the Pfaffian or Pfaffian-type method \cite{RN207, RN480, RN265, RN218, RN269}, which shows significant usefulness in solving statistical lattice models \cite{RN397}. There are also solutions neither from the transfer matrix method nor from the combinatorial formalism, e.g., from the anticommuting variable integrals \cite{RN209} and the Grassmann integrals \cite{RN674}. Among all the derivations of the solution, the Schultz-Mattis-Lieb (SML) method \cite{RN76} using the fermionic representation of the transfer matrix is of particular interest to us, as it has considerably simplified and clarified Onsager's original algebraic derivation.

This paper aims to employ the SML method to derive the partition function of a finite square lattice Ising model under the Brascamp-Kunz (B-K) BCs \cite{RN410}. The introduction of B-K BCs is a breakthrough in the study of Fisher zeros of Ising model \cite{RN308}. Kaufman \cite{RN73} first showed that the finite lattice partition function under the toroidal BCs consists of the sum of four product terms, which makes the calculation of Fisher zeros analytically intractable. While under the B-K BCs, the finite lattice partition function can be expressed in a double product form, such that the Fisher zeros can be solved explicitly and found to lie precisely on well-defined loci. The use of B-K BCs enables a rigorous determination of the Fisher loci and the density distribution function of Fisher zeros \cite{RN411} in the thermodynamic limit. Since the B-K BCs were introduced, the property of the partition function under these BCs, as well as the method of deriving it, has attracted significant attention. In the case of isotropic interactions, the partition function of the model was first obtained via its dual system \cite{RN410}, which had been exactly solved before \cite{RN459}. The solution in the case of anisotropic interactions was also proposed \cite{RN474, RN432, RN455}. A recent work by the authors of the present paper rediscovered the result with isotropic interactions using a mapping into the special BCs for the free-fermion model \cite{RN671}. The existing derivations of the partition function under the B-K BCs are within the framework of Pfaffian-type method. However, we have found no transfer-matrix-based derivation. In this work we contribute to this field by providing such a derivation using the SML method. In addition to the transfer matrix approach related to the toroidal BCs \cite{RN73, RN76, RN666, RN668}, there have been transfer-matrix-based studies under various alternative BCs \cite{RN638, RN474, RN659, RN508, RN637}. Our work will introduce a member to this family. 

The remainder of this paper is organized as follows. In Sec.~\ref{model} the square lattice Ising model under the B-K BCs is introduced. In Sec.~\ref{derive} we give the derivation using the SML method in detail, and present the partition function of a finite lattice. To deal with the difference between the B-K and toroidal BCs, we adopt a technique of taking certain limit of the interactions on the boundaries. Discussion and summary are given in Sec.~\ref{discuss}.

\section{Model} \label{model}
\begin{figure}[h]
\centering
\includegraphics{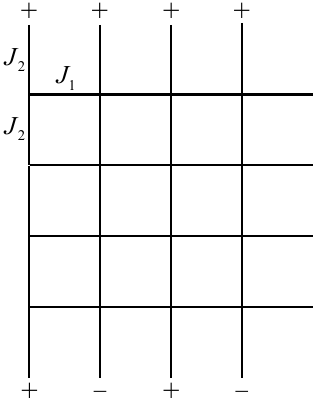}
\caption{The Ising model under the B-K BCs on the $4\times4$ lattice. The interactions in the horizontal and vertical directions are marked.} \label{fig1}
\end{figure}

The system we study is an Ising model on a square lattice of $M$ rows and $2N$ columns, with nearest-neighbour interactions and in the absence of a magnetic field. Each spin $s_i$ of the model can take two values $\pm 1$. The Hamiltonian is given by
\begin{equation}
H = - J_1\sum\limits_{\left\langle {i,j} \right\rangle_{\parallel}} {s_i}{s_j} - J_2\sum\limits_{\left\langle {i,j} \right\rangle_\bot} {s_i}{s_j}~,  \label{eq1}
\end{equation}
where $\left\langle {i,j} \right\rangle_{\parallel}$ and $\left\langle {i,j} \right\rangle_\bot$ denote the nearest neighbours in the horizontal and vertical directions, respectively, and $J_1$ and $J_2$ are the corresponding interaction constants. The system is set under the B-K BCs \cite{RN410}, which can be described as follows: periodic BCs in the $N$ direction, $2N$ ``$+$'' spins on the upper edge (the zeroth row), and $2N$ alternating spins ``$+- \cdots +-$'' on the lower edge [the ($M+1$)th row]. That is, the system is arranged on a cylindrical surface with fixed upper and lower boundaries. Figure \ref{fig1} shows an example on the $4\times4$ lattice.

The partition function of the system is defined as the sum of the Boltzmann factors over all possible spin configurations
\begin{equation}
Z = \sum\limits_{\left\{ s_i \right\} = \pm 1} e^{-\beta H\left( \left\{ s_i \right\} \right)}~,  \label{eq2} 
\end{equation}
with $\beta=1/k_BT$. We aim to find the mathematically exact solution of the partition function under the B-K BCs, using an approach different from those in the previous studies. In Sec.~\ref{derive} we give the detailed derivation.

\section{Derivation and Result} \label{derive}
\begin{figure}[h]
\centering
\includegraphics{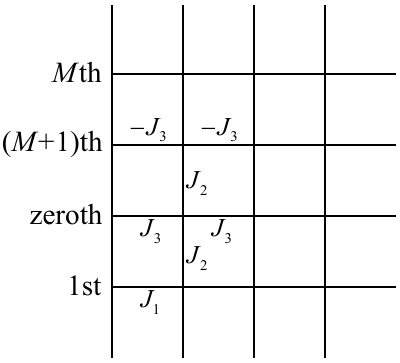}
\caption{The new system under the toroidal BCs, with the interactions marked. The horizontal and vertical interaction constants are given in Eq.~\eqref{eq1}, except that those in the zeroth and ($M+1$)th rows are $J_3$ and $-J_3$, respectively.} \label{fig2}
\end{figure}

The Ising model under the toroidal BCs is the most commonly studied case. The SML method using the row-to-row transfer matrix is originally applied to the case of toroidal BCs \cite{RN76}. To employ the SML method to derive the solution under the B-K BCs, we use a technique of taking certain limit of the interactions on the upper and lower boundaries. Consider a system on the $(M+2)\times2N$ lattice under the toroidal BCs, with the same interaction constants as given in Eq.~\eqref{eq1} except for those in the zeroth and ($M+1$)th rows. Set the interaction constant in the zeroth row as $J_3$ and that in the ($M+1$)th row as $-J_3$, as shown in Fig.~\ref{fig2}. Note that the spins in the zeroth and ($M+1$)th rows now are not fixed. Denote the partition function of this system by $Z(J_3)$. Then we consider the limit when $J_3 \to + \infty$
\begin{equation}
\mathop {\lim}\limits_{J_3 \to + \infty} \frac{1}{e^{4N\beta J_3}}Z(J_3)~.  \label{eq3}
\end{equation}
Notice that 
\begin{align*}
&\mathop {\lim}\limits_{J_3 \to + \infty} \frac{1}{e^{\beta J_3}} e^{\beta J_3 s_1 s_2} = \left\{ \begin{array}{*{20}{l}}{1, s_1 s_2 = 1}\\{0, s_1 s_2 = -1}\end{array} \right. , \\ 
&\mathop {\lim}\limits_{J_3 \to + \infty} \frac{1}{e^{\beta J_3}} e^{-\beta J_3 s_1 s_2} = \left\{ \begin{array}{*{20}{l}}{0, s_1 s_2 = 1}\\{1, s_1 s_2 = -1}\end{array} \right. .
\end{align*}
We can verify that only two choices for the spin configurations in the zeroth row---``$++\cdots++$'' and ``$--\cdots--$''---can make a non-zero contribution to the partition function in the limit of Eq.~\eqref{eq3}. Similarly, we have only two choices for the spin configurations in the ($M+1$)th row---``$+-\cdots+-$'' and ``$-+\cdots-+$''. Therefore, we obtain four sets of BCs surviving in Eq.~\eqref{eq3}: \\
(\romannumeral1) zeroth: $++\cdots++$; ($M+1$)th: $+-\cdots+-$ (B-K BCs)\\
(\romannumeral2) zeroth: $++\cdots++$; ($M+1$)th: $-+\cdots-+$\\
(\romannumeral3) zeroth: $--\cdots--$; ($M+1$)th: $+-\cdots+-$\\
(\romannumeral4) zeroth: $--\cdots--$; ($M+1$)th: $-+\cdots-+$.

We demonstrate that the partition functions under these four sets of BCs are identical. Firstly, it is straightforward to verify $Z_{(\rm{\romannumeral1})}=Z_{(\rm{\romannumeral2})}$ and $Z_{(\rm{\romannumeral3})}=Z_{(\rm{\romannumeral4})}$ by noticing that there are periodic BCs along the $N$ direction. Secondly, since the Hamiltonian is conserved when all spins are reversed, we have $Z_{(\rm{\romannumeral1})}=Z_{(\rm{\romannumeral4})}$ and $Z_{(\rm{\romannumeral2})}=Z_{(\rm{\romannumeral3})}$. Now we can see that the partition function under the B-K BCs is transformed into a quarter of the partition function under the toroidal BCs in a certain limit, i.e., 
\begin{equation}
Z_{{\rm B}\textit{-}{\rm K}} = \frac{1}{4} \mathop{\lim}\limits_{J_3 \to + \infty} \frac{1}{e^{4N\beta J_3}}Z(J_3)~.  \label{eq4}
\end{equation}
It is natural to express $Z(J_3)$ in the transfer matrix formalism, and the SML method can then be employed. 

\subsection{Transfer matrix} \label{trans}
The row-to-row transfer matrix has been defined and studied for a long time. The Ising problem can be conveniently formulated using the language of transfer matrix. To do this, we should first use a row on the lattice as a transfer unit to calculate the partition function. Denote the spin configurations in the $j$th row by $\left\{ s_i \right\}_j$, which consist of the states of $2N$ spins $\left\{ (s_1)_j, ... , (s_{2N})_j \right\}$. Define three $2^{2N} \times 2^{2N}$ matrices 
\begin{align}
&V_1 \left(\left\{ s_i \right\}_j,\left\{ s_i \right\}_j\right) = e^{K_1\sum\limits_{i=1}^{2N} (s_i)_j (s_{i+1})_j } \nonumber \\
&V^{\prime}_1 \left(\left\{ s_i \right\}_j,\left\{ s_i \right\}_j\right) = e^{K_3\sum\limits_{i=1}^{2N} (s_i)_j (s_{i+1})_j } \nonumber \\
&V_2 \left(\left\{ s_i \right\}_j,\left\{ s_i \right\}_{j+1}\right) = e^{K_2\sum\limits_{i=1}^{2N} (s_i)_j (s_i)_{j+1} } \label{eq5}
\end{align}
with $K_l = \beta J_l~(l=1,2,3)$ and $(s_{2N+1})_j = (s_1)_j$. Diagonal matrices $V_1$ and $V^{\prime}_1$ take into account the Boltamann factors associated with the interactions within each row, while matrix $V_2$ corresponds to those associated with the interactions between neighbouring rows. According to Eq.~\eqref{eq2}, $Z(J_3)$ in our case can be expressed as 
\begin{align}
&Z(J_3) = \sum\limits_{\left\{ s_i \right\}_1 = \pm 1} \cdots \sum\limits_{\left\{ s_i \right\}_{M+1} = \pm 1} \sum\limits_{\left\{ s_i \right\}_0 = \pm 1}   \nonumber \\
&~~~~~~~~~~~~V_1\left( \left\{ s_i \right\}_1, \left\{ s_i \right\}_1 \right)  V_2\left( \left\{ s_i \right\}_1, \left\{ s_i \right\}_2 \right) \times \cdots    \nonumber \\
&~~~~~~~\times (V^{\prime}_1)^{-1}\left( \left\{ s_i \right\}_{M+1}, \left\{ s_i \right\}_{M+1} \right)  V_2\left( \left\{ s_i \right\}_{M+1}, \left\{ s_i \right\}_0 \right)   \nonumber \\ 
&~~~~~~~\times V^{\prime}_1\left( \left\{ s_i \right\}_0, \left\{ s_i \right\}_0 \right) V_2\left( \left\{ s_i \right\}_0,\left\{ s_i \right\}_1 \right)   \nonumber \\
& = \sum\limits_{\left\{ s_i \right\}_1 = \pm 1} \cdots \sum\limits_{\left\{ s_i \right\}_{M+1} = \pm 1} \sum\limits_{\left\{ s_i \right\}_0 = \pm 1} (V_1 V_2)\left( \left\{ s_i \right\}_1, \left\{ s_i \right\}_2 \right) \times    \nonumber \\
&\cdots \times [(V^{\prime}_1)^{-1} V_2]\left( \left\{ s_i \right\}_{M+1}, \left\{ s_i \right\}_0 \right) \times (V^{\prime}_1 V_2)\left( \left\{ s_i \right\}_0,\left\{ s_i \right\}_1 \right)   \nonumber \\
& = \mathrm{Tr} \left[ (V_1 V_2)^M (V^{\prime}_1)^{-1} V_2 V^{\prime}_1 V_2 \right].  \label{eq6}
\end{align}
The partition function is now written as the trace of a $2^{2N} \times 2^{2N}$ matrix. Equation \eqref{eq4} then becomes 
\begin{equation}
Z_{{\rm B}\textit{-}{\rm K}} = \frac{1}{4} \mathop{\lim}\limits_{K_3 \to +\infty} \frac{1}{e^{4NK_3}} \mathrm{Tr} \left[ (V_1 V_2)^M (V^{\prime}_1)^{-1} V_2 V^{\prime}_1 V_2 \right].   \label{eq7}
\end{equation}

The explicit expressions of $V_1$, $V^{\prime}_1$ and $V_2$ are clearly given in the Pauli representation \cite{RN72, RN73, RN76}. Use the notations
\begin{align}
&\sigma_i^z = {\bf{1}} \otimes \cdots \otimes {\bf{1}} \otimes \underbrace {\sigma^z}_{i \rm{th}} \otimes {\bf{1}} \otimes \cdots \otimes {\bf{1}},  \nonumber \\
&\sigma_i^x = {\bf{1}} \otimes \cdots \otimes {\bf{1}} \otimes \underbrace {\sigma^x}_{i \rm{th}} \otimes {\bf{1}} \otimes \cdots \otimes {\bf{1}},  \label{eq8}
\end{align}
where $\bf{1}$ is the identity matrix and $\sigma^z$ and $\sigma^x$ are two of the Pauli matrices
\begin{equation}
\sigma^z = \left( \begin{array}{*{20}{c}} 1&0\\0&{-1} \end{array} \right),~ \sigma^x = \left( \begin{array}{*{20}{c}} 0&1\\1&0\end{array} \right),~ \sigma^y = \left( \begin{array}{*{20}{c}} 0&-i\\i&0\end{array} \right).  \label{eq9}
\end{equation}
$V_1$, $V^{\prime}_1$ and $V_2$ can be given explicitly
\begin{align}
&V_1 = e^{K_1 (\sigma_1^z \sigma_2^z + \cdots + \sigma_{2N - 1}^z \sigma_{2N}^z + \sigma_{2N}^z \sigma_1^z)},  \nonumber \\
&V^{\prime}_1 = e^{K_3 (\sigma_1^z \sigma_2^z + \cdots + \sigma_{2N - 1}^z \sigma_{2N}^z + \sigma_{2N}^z \sigma_1^z)},  \nonumber \\
&V_2 = (2\sinh 2K_2)^N e^{K_2^* (\sigma_1^x + \cdots + \sigma_{2N}^x)},  \label{eq10}
\end{align}
with $K_2^*$ defined as
\begin{subequations}
\begin{equation}
\tanh K_2^* = e^{-2 K_2}  \label{eq11a}
\end{equation}
or equivalently
\begin{equation}
\sinh 2K_2 \sinh 2K_2^* = 1.  \label{eq11b}
\end{equation}
\end{subequations}
Note that in this representation the matrices have the same index of rows and of columns (i.e., the same order of spin configurations appearing in rows and in columns). From Eq.~\eqref{eq10} it is clear to verify that, the transfer matrix of a two-dimensional classical Ising model is mapped into the density operator of a one-dimensional quantum Ising model.

As suggested by Ref.~\cite{RN76}, we perform a canonical transformation $\sigma^x \to \sigma^z$, $\sigma^z \to -\sigma^x$ which keeps the commutation rules of Pauli matrices invariant. Under this transformation $V_1$, $V^{\prime}_1$ and $V_2$ is written as
\begin{align}
&V_1 = e^{K_1 (\sigma_1^x \sigma_2^x + \cdots + \sigma_{2N - 1}^x \sigma_{2N}^x + \sigma_{2N}^x \sigma_1^x)},  \nonumber \\
&V^{\prime}_1 = e^{K_3 (\sigma_1^x \sigma_2^x + \cdots + \sigma_{2N - 1}^x \sigma_{2N}^x + \sigma_{2N}^x \sigma_1^x)},  \nonumber \\
&V_2 = (2\sinh 2K_2)^N e^{K_2^* (\sigma_1^z + \cdots + \sigma_{2N}^z)},  \label{eq12}
\end{align}
and Eq.~\eqref{eq7} still holds. Now the transfer matrices can be chosen to be symmetric
\begin{align}
&V = V_2^{1/2} V_1 V_2^{1/2},~V^{\prime\prime} = V_2^{1/2} (V^{\prime}_1)^{-1} V_2^{1/2},  \nonumber \\
&V^{\prime} = V_2^{1/2} V^{\prime}_1 V_2^{1/2},  \label{eq13}
\end{align}
and the partition function in Eq.~\eqref{eq7} is expressed using these matrices
\begin{equation}
Z_{{\rm B}\textit{-}{\rm K}} = \frac{1}{4} \mathop{\lim}\limits_{K_3 \to +\infty} \frac{1}{e^{4NK_3}} \mathrm{Tr} \left[ V^M V^{\prime\prime} V^{\prime} \right].  \label{eq14}
\end{equation}
The fermionic representation will be introduced based on Eq.~\eqref{eq12} in the next section.

\subsection{Fermionic representation} \label{fermi}
\subsubsection{Jordan-Wigner transformation}  \label{j-w}
Define the spin raising and lowering operators
\begin{equation}
\sigma_i^+ = \frac{1}{2}(\sigma_i^x + i\sigma_i^y),~\sigma_i^- = \frac{1}{2}(\sigma_i^x - i\sigma_i^y)~.  \label{eq15}
\end{equation}
It is straightforward to find
\begin{equation}
\sigma_i^x = \sigma_i^+ + \sigma_i^-,~\sigma_i^z = 2\sigma_i^+ \sigma_i^- - {\bf{1}}   \label{eq16}
\end{equation}
(here ${\bf{1}}$ is a $2^{2N}\times2^{2N}$ identity matrix). Equation \eqref{eq12} is then written as
\begin{align}
&\!\!\!\!V_1 = e^{K_1 [ (\sigma_1^+ + \sigma_1^-) (\sigma_2^+ + \sigma_2^-) + \cdots + (\sigma_{2N}^+ + \sigma_{2N}^-) (\sigma_1^+ + \sigma_1^-) ]},  \nonumber \\
&\!\!\!\!V^{\prime}_1 = e^{K_3 [ (\sigma_1^+ + \sigma_1^-) (\sigma_2^+ + \sigma_2^-) + \cdots + (\sigma_{2N}^+ + \sigma_{2N}^-) (\sigma_1^+ + \sigma_1^-) ]},  \nonumber \\
&\!\!\!\!V_2 = (2\sinh 2K_2)^N e^{K_2^* [ (2\sigma_1^+ \sigma_1^- - {\bf{1}}) + \cdots + (2\sigma_{2N}^+ \sigma_{2N}^- - {\bf{1}}) ]}.  \label{eq17}
\end{align}
Now we employ the SML method and use the Jordan-Wigner transformation \cite{RN497} to introduce the fermion annihilation and creation operators
\begin{align}
&C_m = e^{i\pi \sum\limits_{i=1}^{m-1} {\sigma_i^+ \sigma_i^-}}\sigma_m^-,~ C_m^\dagger = e^{i\pi \sum\limits_{i=1}^{m-1} {\sigma_i^+ \sigma_i^-}}\sigma_m^+,  \nonumber \\
&m = 1, \cdots, 2N.  \label{eq18}
\end{align}
The anticommutation rules of fermion operators can be examined
\begin{equation}
\left\{ C_m, C_n \right\}=0,~ \left\{ C_m^\dagger, C_n^\dagger \right\}=0,~ \left\{ C_m, C_n^\dagger \right\}=\delta_{mn}~.  \label{eq19}
\end{equation}
It is straightforward to verify $e^{2i\pi \sum\limits_{i=1}^{m-1} {\sigma_i^+ \sigma_i^-}}={\bf{1}}$ and
\begin{equation}
\sigma_m^+ \sigma_m = C_m^\dagger C_m~,  \label{eq20}
\end{equation}
so that the inverse transformation is given by
\begin{align}
&\sigma_m^- = e^{i\pi \sum\limits_{i=1}^{m-1} {C_i^\dagger C_i}}C_m,~ \sigma_m^+ = e^{i\pi \sum\limits_{i=1}^{m-1} {C_i^\dagger C_i}}C_m^\dagger,   \nonumber \\
&m = 1, \cdots, 2N.  \label{eq21}
\end{align}

To translate the Pauli representation into the fermionic representation, we need to do some algebra. Using the fermion anticommutation relations [Eq.~\eqref{eq19}] we can identify
\begin{align}
&C_m e^{i\pi C_m^\dagger C_m} = -C_m,~ e^{i\pi C_m^\dagger C_m} C_m = C_m,  \nonumber \\
&C_m^\dagger e^{i\pi C_m^\dagger C_m} = C_m^\dagger,~ e^{i\pi C_m^\dagger C_m} C_m^\dagger = -C_m^\dagger .   \label{eq22}
\end{align}
Then we can derive the following identities from Eqs.~\eqref{eq21} and \eqref{eq22}
\begin{align}
&\sigma_m^+ \sigma_{m+1}^+ = C_m^\dagger C_{m+1}^\dagger,~ \sigma_m^+ \sigma_{m+1}^- = C_m^\dagger C_{m+1},  \nonumber \\
&\sigma_m^- \sigma_{m+1}^+ = -C_m C_{m+1}^\dagger,~ \sigma_m^- \sigma_{m+1}^- = -C_m C_{m+1}.   \label{eq23}
\end{align}
Now $V_1$, $V^{\prime}_1$ and $V_2$ are represented by the fermion operators by substituting Eqs.~\eqref{eq20} and \eqref{eq23} into Eqs.~\eqref{eq17}
\begin{align}
&V_1 = e^{K_1 \left[ \sum\limits_{j=1}^{2N-1} (C_j^\dagger - C_j) (C_{j+1}^\dagger + C_{j+1}) - U(C_{2N}^\dagger - C_{2N}) (C_1^\dagger + C_1) \right]},  \nonumber \\
&V^{\prime}_1 = e^{K_3 \left[ \sum\limits_{j=1}^{2N-1} (C_j^\dagger - C_j) (C_{j+1}^\dagger + C_{j+1}) - U(C_{2N}^\dagger - C_{2N}) (C_1^\dagger + C_1) \right]},  \nonumber \\
&V_2 = (2\sinh 2K_2)^N e^{K_2^* \sum\limits_{j=1}^{2N} (2C_j^\dagger C_j - {\bf{1}})},  \label{eq24}
\end{align}
with
\begin{equation}
U = e^{i\pi \sum\limits_{j=1}^{2N} C_j^\dagger C_j}.  \label{eq25}
\end{equation}

To deal with the difficulty caused by the operator $U$ in Eq.~\eqref{eq24}, we adopt a technique used in Refs.~\cite{RN73, RN668}. Notice that all terms in $V_1$, $V^{\prime}_1$ and $V_2$ involve bilinear products of fermion operators, the evenness or oddness of the total number $\sum\nolimits_{j=1}^{2N} C_j^\dagger C_j$ is conserved \cite{RN76}. Therefore $U$ commutes with each of them
\begin{equation}
\left[U, V_1\right] = \left[U, V^{\prime}_1\right] = \left[U, V_2\right] = 0.  \label{eq26}
\end{equation}
More exactly, $U$ commutes with bilinear products of fermion operators. By using $U^2={\bf{1}}$ we find 
\begin{align}
&~~~~(1+U) \times U(C_{2N}^\dagger - C_{2N})(C_1^\dagger + C_1)   \nonumber \\
& = (1+U) (C_{2N}^\dagger - C_{2N})(C_1^\dagger + C_1),  \nonumber \\
&~~~~(1-U) \times U(C_{2N}^\dagger - C_{2N})(C_1^\dagger + C_1)   \nonumber \\
& = -(1-U) (C_{2N}^\dagger - C_{2N})(C_1^\dagger + C_1).  \label{eq27}
\end{align}
This leads to
\begin{equation}
(1+U)V_1 = (1+U)V_1^+,~ (1-U)V_1 = (1-U)V_1^-,  \label{eq28}
\end{equation}
where
\begin{equation}
V_1^{\pm} = e^{K_1 \left[ \sum\limits_{j=1}^{2N} (C_j^\dagger - C_j) (C_{j+1}^\dagger + C_{j+1}) \right]},  \label{eq29}
\end{equation}
with anticyclic definition of $C_{2N+1}^\dagger$ and $C_{2N+1}$
\begin{equation}
C_{2N+1}^\dagger = -C_1^\dagger,~ C_{2N+1} = -C_1  \label{eq30}
\end{equation}
for $V_1^+$ and cyclic definition
\begin{equation}
C_{2N+1}^\dagger = C_1^\dagger,~ C_{2N+1} = C_1  \label{eq31}
\end{equation}
for $V_1^-$. ${V^{\prime}_1}^{\pm}$ are determined similarly. Now we can express the transfer matrix $V$ in Eq.~\eqref{eq13} by using the projection operators $\frac{1}{2}(1 \pm U)$
\begin{align}
V &= \frac{1}{2}(1+U)V + \frac{1}{2}(1-U)V  \nonumber \\
&= \frac{1}{2}(1+U) V_2^{1/2} V_1^+ V_2^{1/2} + \frac{1}{2}(1-U) V_2^{1/2} V_1^- V_2^{1/2}  \nonumber \\
&\equiv \frac{1}{2}(1+U)V^+ + \frac{1}{2}(1-U)V^-.  \label{eq32}
\end{align}
${V^{\prime\prime}}^{\pm}$ and ${V^{\prime}}^{\pm}$ [see Eq.~\eqref{eq13}] are defined similarly. It is seen that the eigenvectors of $\frac{1}{2}(1+U)V^+$ [also those of $\frac{1}{2}(1+U){V^{\prime\prime}}^+$ and $\frac{1}{2}(1+U){V^{\prime}}^+$] involve even numbers of fermions, and those of $\frac{1}{2}(1-U)V^-$ [also those of $\frac{1}{2}(1-U){V^{\prime\prime}}^-$ and $\frac{1}{2}(1-U){V^{\prime}}^-$] involve odd numbers of fermions. 

The next step is to consider $V^M V^{\prime\prime} V^{\prime}$ in Eq.~\eqref{eq14}. Making use of
\begin{subequations}
\begin{align}
&\left[\frac{1}{2}(1+U)\right]^k = \frac{1}{2}(1+U),~\left[\frac{1}{2}(1-U)\right]^k = \frac{1}{2}(1-U),   \nonumber \\
& k \in {N_+},   \label{eq33a}
\end{align}
and
\begin{equation}
(1+U)(1-U) = 0,  \label{eq33b}
\end{equation}
\end{subequations}
it is clear to show
\begin{align}
V^M V^{\prime\prime} V^{\prime} &= \frac{1}{2}(1+U) (V^+)^M {V^{\prime\prime}}^+ {V^{\prime}}^+   \nonumber \\
&~~~~ + \frac{1}{2}(1-U) (V^-)^M {V^{\prime\prime}}^- {V^{\prime}}^-.  \label{eq34}
\end{align}
Like we have analysed for $V$, $V^M V^{\prime\prime} V^{\prime}$ is divided into two parts, which are referred to as the even and odd parts henceforth. Now the operator $U$ in the terms of $V_1$ and $V^{\prime}_1$ is eliminated, and $V_1^{\pm}$ and ${V^{\prime}_1}^{\pm}$ are introduced. Further transformation in the fermionic representation is based on Eqs.~\eqref{eq29}--\eqref{eq31}. The trace of Eq.~\eqref{eq34} is the sum of four terms. Note that when the horizontal interactions are uniform (the usual toroidal case), the trace is actually Kaufman's solution \cite{RN73}, and the four terms in the trace can be seen as four Pfaffians \cite{RN465}. 

\subsubsection{Direct product decomposition}  \label{dpd}
Via the Jordan-Wigner transformation we obtain the expressions of $V_1^{\pm}$, ${V^{\prime}_1}^{\pm}$ and $V_2$ in terms of fermion operators. It is expected to perform a direct product decomposition to these matrices so that the diagonalization will be simplified. Following the SML method we take the linear canonical transformation to a new set of fermion operators
\begin{align}
&\eta_q = (2N)^{-1/2} e^{i\pi /4} \sum\limits_{j=1}^{2N} e^{-iqj}C_j~,  \nonumber \\
&\eta_q^\dag = (2N)^{-1/2} e^{-i\pi /4} \sum\limits_{j=1}^{2N} e^{iqj}C_j^\dag~.  \label{eq35}
\end{align}
It can be examined that $\eta_q$'s and $\eta_q^\dag$'s obey the fermion anticommutation rules [as in Eq.~\eqref{eq19}]. The inverse transformation is straightforward
\begin{align}
&C_j = (2N)^{-1/2} e^{-i\pi /4} \sum\limits_{q} e^{iqj}\eta_q~,  \nonumber \\
&C_j^\dag = (2N)^{-1/2} e^{i\pi /4} \sum\limits_{q} e^{-iqj}\eta_q^\dag~.  \label{eq36}
\end{align}
The index $q$ is set as
\begin{equation}
q = \pm \frac{(2l-1)\pi}{2N},~l = 1, \cdots, N   \label{eq37}
\end{equation}
for the anticyclic condition [Eq.~\eqref{eq30}] for $V_1^+$ in the even part, and as
\begin{equation}
q = 0,~\pi,~\pm \frac{2l\pi}{2N},~l = 1, \cdots, N-1   \label{eq38}
\end{equation}
for the cyclic condition [Eq.~\eqref{eq31}] for $V_1^-$ in the odd part. With this notations we can also refer to $\eta_q$ as $\eta_{\pm(2l-1)}$ or $\eta_{\pm(2l)}$ without loss of clarity. 

Now we can express $V_1^{\pm}$, ${V^{\prime}_1}^{\pm}$ and $V_2$ in terms of $\eta_q$'s and $\eta_q^\dag$'s. To do this, some algebra is needed. First it is easy to see
\begin{equation}
\sum\limits_{j=1}^{2N} C_j^\dagger C_j = \sum\limits_{q} \eta_q^\dag \eta_q~.   \label{eq39}
\end{equation}
This yields $V_2$ in Eq.~\eqref{eq24} in the form
\begin{equation}
V_2^+ = (2\sinh 2K_2)^N e^{2K_2^* \sum\limits_{l=1}^{N} \left(\eta_{2l-1}^\dag \eta_{2l-1} + \eta_{-(2l-1)}^\dag \eta_{-(2l-1)} - 1\right)}   \label{eq40}
\end{equation}
for the even part, and 
\begin{align}
&\!\!V_2^- = (2\sinh 2K_2)^N  \nonumber \\
&\!\! \times e^{2K_2^* \left[ \sum\limits_{l=1}^{N-1} \left(\eta_{2l}^\dag \eta_{2l} + \eta_{-2l}^\dag \eta_{-2l} - 1\right) + \left(\eta_0^\dag \eta_0 - \frac{1}{2}\right) + \left(\eta_{\pi}^\dag \eta_{\pi} - \frac{1}{2}\right) \right]}   \label{eq41}
\end{align}
for the odd part. Next we calculate
\begin{align}
&\sum\limits_{j=1}^{2N} \left( C_j^\dagger - C_j \right) \left( C_{j+1}^\dagger + C_{j+1} \right)   \nonumber \\
= &\sum\limits_{j=1}^{2N} \left( C_j^\dagger C_{j+1}^\dagger + C_{j+1}C_j + C_j^\dagger C_{j+1} + C_{j+1}^\dagger C_j \right)   \nonumber \\
= &\frac{1}{2N}\sum\limits_{j = 1}^{2N} \sum\limits_q \sum\limits_{q'} \left[ e^{i\frac{\pi}{2}} e^{-i(q+q')j} e^{-iq'} \eta_q^\dag \eta_{q'}^\dag    \right.  \nonumber \\
&~~~~~~~~~~~+ e^{-i\frac{\pi}{2}} e^{i(q+q')j} e^{iq} \eta_q \eta_{q'} + e^{-i(q-q')j} e^{iq'} \eta_q^\dag \eta_{q'}  \nonumber \\
&~~~~~~~~~~~+ \left. e^{-i(q-q')j} e^{-iq} \eta_q^\dag \eta_{q'} \right]   \nonumber \\
= &\sum\limits_q \left[ e^{i\frac{\pi}{2}} e^{iq} \eta_q^\dag \eta_{-q}^\dag + e^{-i\frac{\pi}{2}} e^{iq} \eta_q \eta_{-q} + (e^{iq}+e^{-iq})\eta_q^\dag \eta_q \right]   \nonumber \\
= &\sum\limits_q \left[ 2\cos q\times \eta_q^\dag \eta_q + ie^{iq}\left(\eta_q^\dag \eta_{-q}^\dag - \eta_q \eta_{-q}\right) \right].   \label{eq42}
\end{align} 
For the even part the above result can be written as
\begin{align}
&~\sum\limits_{j=1}^{2N} \left( C_j^\dagger - C_j \right) \left( C_{j+1}^\dagger + C_{j+1} \right)   \nonumber \\
&\!\!\!\!\!= 2\sum\limits_{l=1}^{N} \left[ \cos \frac{(2l-1)\pi}{2N} \left( \eta_{2l-1}^\dag \eta_{2l-1} + \eta_{-(2l-1)}^\dag \eta_{-(2l-1)} \right)  \right.  \nonumber \\
&\!\!\!\!\!+ \left. \sin \frac{(2l-1)\pi}{2N} \left( \eta_{2l-1} \eta_{-(2l-1)} - \eta_{2l-1}^\dag \eta_{-(2l-1)}^\dag \right) \right],   \label{eq43}
\end{align}
while for the odd part it is written as
\begin{align}
&~~~~\sum\limits_{j=1}^{2N} \left( C_j^\dagger - C_j \right) \left( C_{j+1}^\dagger + C_{j+1} \right)   \nonumber \\
&= 2\sum\limits_{l=1}^{N-1} \left[ \cos \frac{2l\pi}{2N} \left( \eta_{2l}^\dag \eta_{2l} + \eta_{-2l}^\dag \eta_{-2l} \right)   \right.  \nonumber \\
& + \left. \sin \frac{2l\pi}{2N} \left( \eta_{2l} \eta_{-2l} - \eta_{2l}^\dag \eta_{-2l}^\dag \right) \right] + 2\eta_0^\dag \eta_0 - 2\eta_{\pi}^\dag \eta_{\pi}.   \label{eq44}
\end{align}
This yields $V_1^{\pm}$ in Eq.~\eqref{eq29} in the form
\begin{widetext}
\begin{equation}
V_1^+ = e^{2K_1 \sum\limits_{l=1}^{N} \left[ \cos \frac{(2l-1)\pi}{2N} \left( \eta_{2l-1}^\dag \eta_{2l-1} + \eta_{-(2l-1)}^\dag \eta_{-(2l-1)} \right) + \sin \frac{(2l-1)\pi}{2N} \left( \eta_{2l-1} \eta_{-(2l-1)} - \eta_{2l-1}^\dag \eta_{-(2l-1)}^\dag \right) \right]}  \label{eq45}
\end{equation}
and
\begin{equation}
V_1^- = e^{2K_1 \left\{ \sum\limits_{l=1}^{N-1} \left[ \cos \frac{2l\pi}{2N} \left( \eta_{2l}^\dag \eta_{2l} + \eta_{-2l}^\dag \eta_{-2l} \right) + \sin \frac{2l\pi}{2N} \left( \eta_{2l} \eta_{-2l} - \eta_{2l}^\dag \eta_{-2l}^\dag \right) \right] + \eta_0^\dag \eta_0 - \eta_{\pi}^\dag \eta_{\pi} \right\}}.  \label{eq46}
\end{equation}
\end{widetext}
${V_1^{\prime}}^{\pm}$ are given similarly.

Now we have succeeded in constructing a direct product decomposition for the transfer matrices from Eqs.~\eqref{eq40}, \eqref{eq41}, \eqref{eq45} and \eqref{eq46}: 
\begin{equation}
V_1^+ = \mathop{\prod\nolimits_\otimes}\limits_{l=1}^N V_{1,l}^+,~ V_1^- = \left( \mathop{\prod\nolimits_\otimes}\limits_{l=1}^{N-1} V_{1,l}^- \right) \otimes V_{1,0}^- \otimes V_{1,\pi}^-   \label{eq47}
\end{equation}
with
\begin{widetext}
\begin{align}
&V_{1,l}^+ = e^{2K_1 \left[ \cos \frac{(2l-1)\pi}{2N} \left( \eta_{2l-1}^\dag \eta_{2l-1} + \eta_{-(2l-1)}^\dag \eta_{-(2l-1)} \right) + \sin \frac{(2l-1)\pi}{2N} \left( \eta_{2l-1} \eta_{-(2l-1)} - \eta_{2l-1}^\dag \eta_{-(2l-1)}^\dag \right) \right]},   \nonumber \\
&V_{1,l}^- = e^{2K_1 \left[ \cos \frac{2l\pi}{2N} \left( \eta_{2l}^\dag \eta_{2l} + \eta_{-2l}^\dag \eta_{-2l} \right) + \sin \frac{2l\pi}{2N} \left( \eta_{2l} \eta_{-2l} - \eta_{2l}^\dag \eta_{-2l}^\dag \right) \right]},  \nonumber \\
&V_{1,0}^- = e^{2K_1 \eta_0^\dag \eta_0},~ V_{1,\pi}^- = e^{-2K_1 \eta_{\pi}^\dag \eta_{\pi}};  \label{eq48}
\end{align}
and
\end{widetext}
\begin{align}
&V_2^+ = (2\sinh 2K_2)^N \mathop{\prod\nolimits_\otimes}\limits_{l=1}^N V_{2,l}^+,   \nonumber \\
&V_2^- = (2\sinh 2K_2)^N \left( \mathop{\prod\nolimits_\otimes}\limits_{l=1}^{N-1} V_{2,l}^- \right) \otimes V_{2,0}^- \otimes V_{2,\pi}^-   \label{eq49}
\end{align}
with
\begin{align}
&V_{2,l}^+ = e^{2K_2^* \left(\eta_{2l-1}^\dag \eta_{2l-1} + \eta_{-(2l-1)}^\dag \eta_{-(2l-1)} - 1\right)},   \nonumber \\
&V_{2,l}^- = e^{2K_2^* \left(\eta_{2l}^\dag \eta_{2l} + \eta_{-2l}^\dag \eta_{-2l} - 1\right)},   \nonumber \\
&V_{2,0}^- = e^{2K_2^* \left( \eta_0^\dag \eta_0 - \frac{1}{2} \right)},~ V_{2,\pi}^- = e^{2K_2^* \left( \eta_{\pi}^\dag \eta_{\pi} - \frac{1}{2} \right)}~.  \label{eq50}
\end{align}
$V_{1,l}^{\prime\prime \pm}$ and $V_{1,l}^{\prime \pm}$ are defined similarly. Obviously the operator $U$ has the form
\begin{align}
&U^+ = \mathop{\prod\nolimits_\otimes}\limits_{l=1}^N e^{i\pi \left(\eta_{2l-1}^\dag \eta_{2l-1} + \eta_{-(2l-1)}^\dag \eta_{-(2l-1)}\right)},\rm{for}~\rm{the}~\rm{even}~\rm{part};        \nonumber \\
&U^- = \left[ \mathop{\prod\nolimits_\otimes}\limits_{l=1}^{N-1} e^{i\pi \left(\eta_{2l}^\dag \eta_{2l} + \eta_{-2l}^\dag \eta_{-2l}\right)} \right] \otimes e^{i\pi \eta_0^\dag \eta_0} \otimes e^{i\pi \eta_{\pi}^\dag \eta_{\pi}},   \nonumber \\
&~~~~~~~~~~~~~~~~~~~~~~~~~~~~~~~~~~~~~~~~~\rm{for}~\rm{the}~\rm{odd}~\rm{part}.  \label{eq51}
\end{align}
Then we can return to Eq.~\eqref{eq34}. The even and odd parts of $V^M V^{\prime\prime} V^{\prime}$ can be given in terms of direct products respectively:
\begin{align}
&\frac{1}{2}(1+U^+) (V^+)^M {V^{\prime\prime}}^+ {V^{\prime}}^+ = \frac{1}{2} (2\sinh 2K_2)^{(M+2)N}  \nonumber \\
&\times \left\{ \mathop{\prod\nolimits_\otimes}\limits_{l=1}^N {\bar V}_l^+ + \mathop{\prod\nolimits_\otimes}\limits_{l=1}^N e^{i\pi \left(\eta_{2l-1}^\dag \eta_{2l-1} + \eta_{-(2l-1)}^\dag \eta_{-(2l-1)}\right)} {\bar V}_l^+ \right\} \label{eq52}
\end{align}
with
\begin{align}
&{\bar V}_l^+ = \left[(V_{2,l}^+)^{1/2} V_{1,l}^+ (V_{2,l}^+)^{1/2}\right]^M \left[(V_{2,l}^+)^{1/2} V_{1,l}^{\prime\prime+} (V_{2,l}^+)^{1/2}\right]   \nonumber \\
&~~~~~~~~~\times \left[(V_{2,l}^+)^{1/2} V_{1,l}^{\prime+} (V_{2,l}^+)^{1/2}\right];   \label{eq53}
\end{align}
and
\begin{align}
&\frac{1}{2}(1-U^-) (V^-)^M {V^{\prime\prime}}^- {V^{\prime}}^- = \frac{1}{2} (2\sinh 2K_2)^{(M+2)N}   \nonumber \\
&\times \left\{ \left( \mathop{\prod\nolimits_\otimes}\limits_{l=1}^{N-1} {\bar V}_l^- \right) \otimes {\bar V}_0^- \otimes {\bar V}_{\pi}^- \right.  \nonumber \\
&\!\!\!- \left. \left[ \mathop{\prod\nolimits_\otimes}\limits_{l=1}^{N-1} e^{i\pi \left(\eta_{2l}^\dag \eta_{2l} + \eta_{-2l}^\dag \eta_{-2l}\right)} {\bar V}_l^- \right] \otimes e^{i\pi \eta_0^\dag \eta_0}{\bar V}_0^- \otimes e^{i\pi \eta_{\pi}^\dag \eta_{\pi}}{\bar V}_{\pi}^- \right\}   \label{eq54}
\end{align}
with
\begin{align}
&{\bar V}_l^- = \left[(V_{2,l}^-)^{1/2} V_{1,l}^- (V_{2,l}^-)^{1/2}\right]^M \left[(V_{2,l}^-)^{1/2} V_{1,l}^{\prime\prime-} (V_{2,l}^-)^{1/2}\right]   \nonumber \\
&~~~~~~~~~\times \left[(V_{2,l}^-)^{1/2} V_{1,l}^{\prime-} (V_{2,l}^-)^{1/2}\right],  \nonumber \\ 
&{\bar V}_0^- = \left[(V_{2,0}^-)^{1/2} V_{1,0}^- (V_{2,0}^-)^{1/2}\right]^M \left[(V_{2,0}^-)^{1/2} V_{1,0}^{\prime\prime-} (V_{2,0}^-)^{1/2}\right]   \nonumber \\
&~~~~~~~~~\times \left[(V_{2,0}^-)^{1/2} V_{1,0}^{\prime-} (V_{2,0}^-)^{1/2}\right], \nonumber \\
&{\bar V}_{\pi}^- = \left[(V_{2,\pi}^-)^{1/2} V_{1,\pi}^- (V_{2,\pi}^-)^{1/2}\right]^M \left[(V_{2,\pi}^-)^{1/2} V_{1,\pi}^{\prime\prime-} (V_{2,\pi}^-)^{1/2}\right]   \nonumber \\
&~~~~~~~~~\times \left[(V_{2,\pi}^-)^{1/2} V_{1,\pi}^{\prime-} (V_{2,\pi}^-)^{1/2}\right].  \label{eq55}
\end{align}
In the next section, the partition function will be derived based on the direct product decomposition [Eqs.~\eqref{eq52}--\eqref{eq55}].

\subsection{Partition function} \label{solution}
The partition function $Z_{{\rm B}\textit{-}{\rm K}}$ is expressed in Eq.~\eqref{eq14} as the trace of $V^M V^{\prime} V^{\prime\prime}$ in the limit $\lim_{{K_3} \to + \infty} \frac{1}{e^{4NK_3}}$. We have already obtained the $\left\{ \eta \right\}$ fermion representation and the form of direct product decomposition. Substituting Eqs.~\eqref{eq52} and \eqref{eq54} into Eq.~\eqref{eq14} yields
\begin{align}
&\!\!\!\!Z_{{\rm B}\textit{-}{\rm K}} = \frac{1}{8}(2\sinh 2K_2)^{(M+2)N} \left\{ \prod\limits_{l=1}^N \textcircled{1}_l + \prod\limits_{l=1}^N \textcircled{2}_l   \right.  \nonumber \\
&\!\!\!\!\!+ \left. \left(\prod\limits_{l=1}^{N-1} \textcircled{3}_l\right) \times \textcircled{4} \times \textcircled{5}- \left(\prod\limits_{l=1}^{N-1} \textcircled{6}_l\right) \times \textcircled{7} \times \textcircled{8} \right\}   \label{eq56}
\end{align}
with
\begin{align}
&\textcircled{1}_l = \mathop{\lim}\limits_{K_3 \to +\infty} \frac{1}{e^{4K_3}} \mathrm{Tr} \left({\bar V}_l^+\right),  \nonumber \\
&\textcircled{2}_l = \mathop{\lim}\limits_{K_3 \to +\infty} \frac{1}{e^{4K_3}} \mathrm{Tr} \left[ e^{i\pi \left(\eta_{2l-1}^\dag \eta_{2l-1} + \eta_{-(2l-1)}^\dag \eta_{-(2l-1)}\right)} {\bar V}_l^+ \right],   \nonumber \\  
&\textcircled{3}_l = \mathop{\lim}\limits_{K_3 \to +\infty} \frac{1}{e^{4K_3}} \mathrm{Tr} \left({\bar V}_l^-\right),   \nonumber \\
&\textcircled{4} = \mathop{\lim}\limits_{K_3 \to +\infty} \frac{1}{e^{2K_3}} \mathrm{Tr}\left({\bar V}_0^-\right),~ \textcircled{5} = \mathop{\lim}\limits_{K_3 \to +\infty} \frac{1}{e^{2K_3}} \mathrm{Tr}\left({\bar V}_\pi^-\right),  \nonumber \\
&\textcircled{6}_l = \mathop{\lim}\limits_{K_3 \to +\infty} \frac{1}{e^{4K_3}} \mathrm{Tr} \left[e^{i\pi \left(\eta_{2l}^\dag \eta_{2l} + \eta_{-2l}^\dag \eta_{-2l}\right)} {\bar V}_l^-\right],  \nonumber \\
&\textcircled{7} = \mathop{\lim}\limits_{K_3 \to +\infty} \frac{1}{e^{2K_3}} \mathrm{Tr}\left(e^{i\pi \eta_0^\dag \eta_0} {\bar V}_0^-\right),  \nonumber \\
&\textcircled{8} = \mathop{\lim}\limits_{K_3 \to +\infty} \frac{1}{e^{2K_3}} \mathrm{Tr}\left(e^{i\pi \eta_{\pi}^\dag \eta_{\pi}} {\bar V}_{\pi}^-\right).   \label{eq57}
\end{align}
Therefore, the problem reduces to the calculation of the traces of the matrices associated with ${\bar V}_l^+$ and ${\bar V}_l^-$, which are defined in Eqs.~\eqref{eq53} and \eqref{eq55}. In expressing $Z_{{\rm B}\textit{-}{\rm K}}$ as Eqs.~\eqref{eq56} and \eqref{eq57} we have partitioned the factor $\frac{1}{e^{4NK_3}}$. Below we will illustrate that this partition is reasonable.

The explicit expressions of all the relevant matrices can be given in the basis $\left| 0_{-(2l-1)} 0_{2l-1} \right\rangle$, $\left| 1_{-(2l-1)} 1_{2l-1} \right\rangle$, $\left| 0_{-(2l-1)} 1_{2l-1} \right\rangle$, $\left| 1_{-(2l-1)} 0_{2l-1} \right\rangle$ or $\left| 0_{-2l} 0_{2l} \right\rangle$, $\left| 1_{-2l} 1_{2l} \right\rangle$, $\left| 0_{-2l} 1_{2l} \right\rangle$, $\left| 1_{-2l} 0_{2l} \right\rangle$, which are denoted as $\left| 00 \right\rangle$, $\left| 11 \right\rangle$, $\left| 01 \right\rangle$, $\left| 10 \right\rangle$ for convenience. We first consider $\textcircled{1}_l$ in the even part. The matrix ${\bar V}_l^+$ is given in Eq.~\eqref{eq53} with $V_{1,l}^+$, $V_{1,l}^{\prime\prime+}$, $V_{1,l}^{\prime+}$ and $V_{2,l}^+$ defined in Eqs.~\eqref{eq48} and \eqref{eq50}. We observe that $\left| 01 \right\rangle$ and $\left| 10 \right\rangle$ are eigenvectors of ${\bar V}_l^+$ with the same eigenvalue, as
\begin{align}
&V_{1,l}^+ \left| 01 \right\rangle ({\rm{or}}~\left| 10 \right\rangle) = e^{2K_1 \cos \frac{(2l-1)\pi}{2N}} \left| 01 \right\rangle (\rm{or}~\left| 10 \right\rangle),   \nonumber \\
&V_{1,l}^{\prime\prime+} \left| 01 \right\rangle ({\rm{or}}~\left| 10 \right\rangle) = e^{-2K_3 \cos \frac{(2l-1)\pi}{2N}} \left| 01 \right\rangle (\rm{or}~\left| 10 \right\rangle),  \nonumber \\
&V_{1,l}^{\prime+} \left| 01 \right\rangle ({\rm{or}}~\left| 10 \right\rangle) = e^{2K_3 \cos \frac{(2l-1)\pi}{2N}} \left| 01 \right\rangle  ({\rm{or}}~\left| 10 \right\rangle),  \nonumber \\
&V_{2,l}^+ \left| 01 \right\rangle ({\rm{or}}~\left| 10 \right\rangle) = \left| 01 \right\rangle ({\rm{or}}~\left| 10 \right\rangle).  \label{eq58}
\end{align}
We see that in the space of $\left| 01 \right\rangle$ and $\left| 10 \right\rangle$, the $2\times2$ matrix is
\begin{equation}
{\bar V}_l^+ = e^{2MK_1 \cos \frac{(2l-1)\pi}{2N}}{\bf{1}}~.   \label{eq59}
\end{equation}
This matrix has zero contribution to $\textcircled{1}_l$ in the limit $\lim_{K_3 \to + \infty} \frac{1}{e^{4K_3}}$ [see Eq.~\eqref{eq57}]. Thus $\textcircled{1}_l$ is determined in the space of $\left| 00 \right\rangle$ and $\left| 11 \right\rangle$. The same analysis also applies to $\textcircled{2}_l$, $\textcircled{3}_l$ and $\textcircled{6}_l$, so that we know
\begin{equation}
\textcircled{1}_l=\textcircled{2}_l,~ \textcircled{3}_l=\textcircled{6}_l.  \label{eq60}
\end{equation}

In the space of $\left| 00 \right\rangle$ and $\left| 11 \right\rangle$, the explicit forms of $V_{1,l}^+$, $V_{1,l}^{\prime\prime+}$, $V_{1,l}^{\prime+}$ and $V_{2,l}^+$ are given as
\begin{widetext}
\begin{align}
&V_{1,l}^+ = e^{2K_1 \cos \frac{(2l-1)\pi}{2N}}\left( \begin{array}{*{20}{c}} \cosh2K_1 + \sinh2K_1\cos\frac{(2l-1)\pi}{2N} & \sinh2K_1\sin\frac{(2l-1)\pi}{2N} \\ \sinh2K_1\sin\frac{(2l-1)\pi}{2N} & \cosh2K_1 - \sinh2K_1\cos\frac{(2l-1)\pi}{2N}\end{array} \right),  \nonumber \\
&V_{1,l}^{\prime\prime+} = e^{-2K_3 \cos \frac{(2l-1)\pi}{2N}}\left( \begin{array}{*{20}{c}} \cosh2K_3 - \sinh2K_3\cos\frac{(2l-1)\pi}{2N} & -\sinh2K_3\sin\frac{(2l-1)\pi}{2N} \\ -\sinh2K_3\sin\frac{(2l-1)\pi}{2N} & \cosh2K_3 + \sinh2K_3\cos\frac{(2l-1)\pi}{2N}\end{array} \right),  \nonumber \\
&V_{1,l}^{\prime+} = e^{2K_3 \cos \frac{(2l-1)\pi}{2N}}\left( \begin{array}{*{20}{c}} \cosh2K_3 + \sinh2K_3\cos\frac{(2l-1)\pi}{2N} & \sinh2K_3\sin\frac{(2l-1)\pi}{2N} \\ \sinh2K_3\sin\frac{(2l-1)\pi}{2N} & \cosh2K_3 - \sinh2K_3\cos\frac{(2l-1)\pi}{2N}\end{array} \right),  \nonumber \\
&V_{2,l}^+ = \left( \begin{array}{*{20}{c}} e^{-2K_2^*}&{}\\{}&e^{2K_2^*}\end{array} \right).   \label{eq61}
\end{align}
\end{widetext}
$V_{2,l}^+$ is obviously diagonal, and the details of determining the elements of $V_{1,l}^+$, $V_{1,l}^{\prime\prime+}$ and $V_{1,l}^{\prime+}$ follow closely those proposed in Ref.~\cite{RN76}. We take the limit of $\left[(V_{2,l}^+)^{1/2} V_{1,l}^{\prime\prime+} (V_{2,l}^+)^{1/2}\right] \left[(V_{2,l}^+)^{1/2} V_{1,l}^{\prime+} (V_{2,l}^+)^{1/2}\right]$
\begin{widetext}
\begin{align}
&~~~\mathop{\lim}\limits_{K_3 \to +\infty} \frac{1}{e^{4K_3}} \left[(V_{2,l}^+)^{1/2} V_{1,l}^{\prime\prime+} (V_{2,l}^+)^{1/2}\right] \left[(V_{2,l}^+)^{1/2} V_{1,l}^{\prime+} (V_{2,l}^+)^{1/2}\right]  \nonumber \\
&= \frac{1}{4}\sin\frac{(2l-1)\pi}{2N} \left( \begin{array}{*{20}{c}} (e^{-4K_2^*}-1)\sin\frac{(2l-1)\pi}{2N} & -2\sinh2K_2^*\left(1-\cos\frac{(2l-1)\pi}{2N}\right) \\ 2\sinh2K_2^*\left(1+\cos\frac{(2l-1)\pi}{2N}\right) & (e^{4K_2^*}-1)\sin\frac{(2l-1)\pi}{2N}\end{array} \right).   \label{eq62}
\end{align}
\end{widetext}
The result indicates that the partition of $\frac{1}{e^{4NK_3}}$ is reasonable. Then we diagonalize $(V_{2,l}^+)^{1/2} V_{1,l}^+ (V_{2,l}^+)^{1/2}$
\begin{equation}
(V_{2,l}^+)^{1/2} V_{1,l}^+ (V_{2,l}^+)^{1/2} = e^{2K_1 \cos \frac{(2l-1)\pi}{2N}} T \left( \begin{array}{*{20}{c}} e^{\epsilon_l}&{}\\{}&e^{-\epsilon_l}\end{array} \right) T^{-1},   \label{eq63}
\end{equation}
where $\epsilon_l$ is the positive root of 
\begin{widetext}
\begin{equation}
\cosh{\epsilon_l} = \cosh2K_1 \cosh2K_2^* - \sinh2K_1 \sinh2K_2^*\cos\frac{(2l-1)\pi}{2N}   \label{eq64}
\end{equation}
and
\begin{align}
&T = \left( \begin{array}{*{20}{c}} \sinh2K_1\sin\frac{(2l-1)\pi}{2N} & \sinh2K_1\sin\frac{(2l-1)\pi}{2N} \\ e^{\epsilon_l}-e^{-2K_2^*}\left(\cosh2K_1 + \sinh2K_1\cos\frac{(2l-1)\pi}{2N}\right) & ~e^{-\epsilon_l}-e^{-2K_2^*}\left(\cosh2K_1 + \sinh2K_1\cos\frac{(2l-1)\pi}{2N}\right)\end{array} \right),  \nonumber \\
&T^{-1} = \frac{1}{\sinh2K_1\sin\frac{(2l-1)\pi}{2N}(e^{-\epsilon_l}-e^{\epsilon_l})} \left( \begin{array}{*{20}{c}} e^{-\epsilon_l}-e^{-2K_2^*}\left(\cosh2K_1 + \sinh2K_1\cos\frac{(2l-1)\pi}{2N}\right) & -\sinh2K_1\sin\frac{(2l-1)\pi}{2N} \\ -e^{\epsilon_l}+e^{-2K_2^*}\left(\cosh2K_1 + \sinh2K_1\cos\frac{(2l-1)\pi}{2N}\right) & \sinh2K_1\sin\frac{(2l-1)\pi}{2N}\end{array} \right).   \label{eq65}
\end{align}
\end{widetext}
Now substituting Eqs.~\eqref{eq62}--\eqref{eq65} into Eq.~\eqref{eq53} and taking the trace in the limit $\lim_{K_3 \to +\infty} \frac{1}{e^{4K_3}}$, we obtain the result of $\textcircled{1}_l$ by elementary algebra
\begin{align}
\textcircled{1}_l& = \mathop{\lim}\limits_{K_3 \to +\infty} \frac{1}{e^{4K_3}} \mathrm{Tr} \left({\bar V}_l^+\right)   \nonumber \\
&= e^{2MK_1 \cos \frac{(2l-1)\pi}{2N}} \times \frac{\sin^2 \frac{(2l-1)\pi}{2N} (2\sinh2K_2^*)^2}{4}   \nonumber \\
&~~~\times \frac{e^{(M+1)\epsilon_l}-e^{-(M+1)\epsilon_l}}{e^{\epsilon_l}-e^{-\epsilon_l}}~.  \label{eq66}
\end{align}
Making use of an identity
\begin{equation}
\gamma^{M+1} - \gamma^{-(M+1)} = (\gamma - \gamma^{-1}) \prod\limits_{j=1}^M \left(\gamma + \gamma^{-1} - 2\cos\frac{j\pi}{M+1} \right)   \label{eq67}
\end{equation}
we have the form of $\textcircled{1}_l$
\begin{align}
\textcircled{1}_l =&~ e^{2MK_1 \cos \frac{(2l-1)\pi}{2N}} \times \frac{\sin^2 \frac{(2l-1)\pi}{2N} (2\sinh2K_2^*)^2}{4}   \nonumber \\
&\times \prod\limits_{j=1}^M 2\left(\cosh2K_1 \cosh2K_2^* - \sinh2K_1 \sinh2K_2^*  \right.  \nonumber \\
&~~~~~~~~~~~\left. \times \cos\frac{(2l-1)\pi}{2N} - \cos\frac{j\pi}{M+1} \right).  \label{eq68}
\end{align} 

We then turn to the odd part. Notice that $V_{1,0}^-$, $V_{2,0}^-$, $V_{1,0}^{\prime\prime-}$, $V_{1,0}^{\prime-}$ and $V_{1,\pi}^-$, $V_{2,\pi}^-$, $V_{1,\pi}^{\prime\prime-}$, $V_{1,\pi}^{\prime-}$ are all diagonal, and
\begin{equation}
V_{1,0}^{\prime\prime-} = V_{1,\pi}^{\prime-} = \left( \begin{array}{*{20}{c}} 1 & {} \\ {} & e^{-2K_3}\end{array} \right),~ V_{1,0}^{\prime-} = V_{1,\pi}^{\prime\prime-} = \left( \begin{array}{*{20}{c}} 1 & {} \\ {} & e^{2K_3}\end{array} \right).   \label{eq69}
\end{equation}
It is clear that $V_0^-$ and $V_\pi^-$ are independent of $K_3$. We immediately verify that
\begin{equation}
\textcircled{4}=\textcircled{5}=\textcircled{7}=\textcircled{8}=0   \label{eq70}
\end{equation}
from Eq.~\eqref{eq57}. Thus the odd part has no contribution to the partition function.

Now we can go to the final step. Only the even part of the transfer matrix contributes to the partition function. Substituting Eqs.~\eqref{eq60}, \eqref{eq68} and \eqref{eq70} into Eq.~\eqref{eq56}, and making use of $\sum\limits_{l=1}^N \cos \frac{(2l-1)\pi}{2N} = 0$, Eq.~\eqref{eq11b} and the identity
\begin{equation}
\prod\limits_{l=1}^N \sin\frac{(2l-1)\pi}{2N} = \frac{1}{2^{N-1}}~,   \label{eq71}
\end{equation}
the final result turns out to be
\begin{widetext}
\begin{align}
Z_{{\rm B}\textit{-}{\rm K}}& = \frac{1}{8}(2\sinh 2K_2)^{(M+2)N} \times 2\prod\limits_{l=1}^N \textcircled{1}_l    \nonumber \\
&= (2\sinh 2K_2)^{MN} \times \prod\limits_{l=1}^N \prod\limits_{j=1}^M 2\left(\cosh2K_1 \cosh2K_2^* - \sinh2K_1 \sinh2K_2^*\cos\frac{(2l-1)\pi}{2N} - \cos\frac{j\pi}{M+1} \right)   \nonumber \\
&= 2^{2MN} \prod\limits_{l=1}^N \prod\limits_{j=1}^M \left(\cosh2K_1 \cosh2K_2 - \sinh2K_1 \cos\frac{(2l-1)\pi}{2N} - \sinh 2K_2 \cos\frac{j\pi}{M+1} \right).   \label{eq72}
\end{align}
\end{widetext}
Now we accomplish our derivation. Our solution is completely consistent with the result previously published \cite{RN410, RN474, RN432, RN455, RN671}.

\section{Discussion and Summary} \label{discuss}
We have succeeded in deriving the solution of the square lattice Ising model under the B-K BCs by the SML method within the transfer matrix formalism. As shown in Eq.~\eqref{eq72}, the partition function has a double product form, which permits an analytic calculation of the Fisher zeros. In the variable $z=\sinh2K_1$, the Fisher zeros are explicitly solved
\begin{widetext}
\begin{align}
z_{lj,1,2} = \frac{ \sinh 2K_2 \cos \frac{(2l - 1)\pi}{2N} \cos \frac{j\pi}{M+1} \pm i\cosh 2K_2 \sqrt {\sinh^2 2K_2 \sin^2 \frac{j\pi}{M+1} + \sin^2 \frac{(2l-1)\pi}{2N}} }{\cosh^2 2K_2 - \cos^2 \frac{(2l-1)\pi}{2N}},~ l=1,\cdots,N;~j=1,\cdots,M.  \label{eq73}
\end{align}
\end{widetext}
For finite $M$ and $N$, the Fisher zeros in the complex $z$ plane do not lie on the real axis. When the system approaches the thermodynamic limit $M,N \to \infty$, the Fisher loci cut the real axis at $z=\frac{1}{\sinh2K_2}$ and $z=-\frac{1}{\sinh2K_2}$. Hence, the system exhibits a phase transition at the critical temperature determined by $\sinh2K_1\sinh2K_2=1$ ($J_1$ and $J_2$ have the same sign) or $\sinh2K_1\sinh2K_2=-1$ ($J_1$ and $J_2$ have opposite signs). The free energy in the thermodynamic limit is directly obtained from Eq.~\eqref{eq72}
\begin{align}
&\mathop{\lim}\limits_{M,N \to \infty} \frac{1}{2MN}\ln Z_{{\rm B}\textit{-}{\rm K}} = \ln2 + \frac{1}{8\pi^2} \int_0^{2\pi} {d\theta} \int_0^{2\pi} {d\phi}  \nonumber \\
&\ln \left(\cosh2K_1 \cosh2K_2 - \sinh2K_1 \cos\theta - \sinh 2K_2 \cos\phi \right),   \label{eq74}
\end{align}
which is Onsager's well-known solution \cite{RN72}. The physical critical point can also be verified from taking the derivative of this solution with respect to $T$.

Two special cases of the system are of interest. The first is the case that $K_2=0$, i.e., $J_2=0$ and the system reduces to the one-dimensional model. The Fisher zeros in Eq.~\eqref{eq73} become
\begin{equation}
z_{l,1,2} = \pm i \frac{1}{\sin \frac{(2l-1)\pi}{2N}},~  l=1,\cdots,N.   \label{eq75}
\end{equation}
This result agrees with the previous studies of one-dimensional Ising model \cite{RN437}. The second case is that the interactions are isotropic, i.e., $J_1=J_2\equiv J$ and $K_1=K_2\equiv K$. In terms of $\bar z = \sinh2K$, the partition function in Eq.~\eqref{eq72} is expressed as
\begin{align}
Z_{{\rm B}\textit{-}{\rm K}} = 2^{2MN} \bar z^{MN} \prod\limits_{l=1}^N \prod\limits_{j=1}^M &\left( \bar z + \bar z^{-1} - \cos\frac{(2l-1)\pi}{2N}   \right.  \nonumber \\
&~~~-\left. \cos\frac{j\pi}{M+1} \right).   \label{eq76}
\end{align}
The Fisher zeros are
\begin{equation}
\bar z_{lj,1,2} = e^{\pm i\theta_{lj}}   \label{eq77}
\end{equation}
with
\begin{equation}
\theta_{lj} = \arccos \left[\frac{1}{2}\left(\cos\frac{(2l-1)\pi}{2N} + \cos\frac{j\pi}{M+1}\right) \right].    \label{eq78}
\end{equation}
For any finite lattice, the Fisher zeros lie on the unit circle in the complex $\bar z$ plane. In the thermodynamic limit the Fisher loci form a continuous unit circle, thus we find the physical critical point at $\bar z=1$ ($J>0$) or $\bar z=-1$ ($J<0$). When we consider the variable $u=e^{2K}$, the Fisher loci form two circles $\left|u \pm 1\right| = \sqrt{2}$, which were first proposed by Fisher \cite{RN308}. The physical critical point in variable $u$ is $u=\sqrt{2}+1$ ($J>0$) or $u=\sqrt{2}-1$ ($J<0$).

Finally we briefly discuss the difference between the transfer matrix approaches to the B-K BCs and to the toroidal BCs. In Sec.~\ref{derive} we transform the system under the B-K BCs into another system under the toroidal BCs, by setting special interactions on the upper and lower boundaries and taking the limit $\lim_{J_3 \to +\infty} \frac{1}{e^{4N\beta J_3}}$. In Eq.~\eqref{eq34} the product of transfer matrices is divided into the even and odd parts. The effect of the limit $\lim_{J_3 \to +\infty} \frac{1}{e^{4N\beta J_3}}$ is that the contribution of the odd part is $0$ and the contribution from two subparts in the even part is equal, so that we achieve the final double product form of the partition function. While under the usual toroidal BCs, i.e., the horizontal interactions are uniform in our model, $V^M V^{\prime\prime} V^{\prime}$ in Eq.~\eqref{eq34} simply becomes $V^{M+2}$ and the solution is the sum of four double products \cite{RN73}. The detail in the transfer matrix leads to the significant difference between the forms of two solutions. Therefore, it is natural to apply the technique of taking certain limit of certain interactions to other lattices or other BCs, such as the Kagom\'e lattice model \cite{RN82} and the square lattice model under the open BCs \cite{RN508}, in the transfer matrix formalism. The transfer matrix approach in the lattice models under various BCs, in particular the SML method using the fermionic representation, merits further investigation.

\begin{acknowledgments}
We thank Prof.~Xiao-Bao Yang for discussions and Prof.~Hong-Ru Ma for his inspiring notes.
This work was supported by Guangdong Provincial Quantum Science Strategic Initiative (Grants No.~GDZX2203001 and No.~GDZX2403001), Shenzhen Fundamental Research Program (Grant No.~JCYJ20240813153139050), National Natural Science Foundation of China (Grant No.~12474489), and Research Funding for Outbound Postdoctoral Fellows in Shenzhen (Grant No.~SZRCXM2401006).
\end{acknowledgments}


\begin{thebibliography}{56}%
\makeatletter
\providecommand \@ifxundefined [1]{%
 \@ifx{#1\undefined}
}%
\providecommand \@ifnum [1]{%
 \ifnum #1\expandafter \@firstoftwo
 \else \expandafter \@secondoftwo
 \fi
}%
\providecommand \@ifx [1]{%
 \ifx #1\expandafter \@firstoftwo
 \else \expandafter \@secondoftwo
 \fi
}%
\providecommand \natexlab [1]{#1}%
\providecommand \enquote  [1]{``#1''}%
\providecommand \bibnamefont  [1]{#1}%
\providecommand \bibfnamefont [1]{#1}%
\providecommand \citenamefont [1]{#1}%
\providecommand \href@noop [0]{\@secondoftwo}%
\providecommand \href [0]{\begingroup \@sanitize@url \@href}%
\providecommand \@href[1]{\@@startlink{#1}\@@href}%
\providecommand \@@href[1]{\endgroup#1\@@endlink}%
\providecommand \@sanitize@url [0]{\catcode `\\12\catcode `\$12\catcode
  `\&12\catcode `\#12\catcode `\^12\catcode `\_12\catcode `\%12\relax}%
\providecommand \@@startlink[1]{}%
\providecommand \@@endlink[0]{}%
\providecommand \url  [0]{\begingroup\@sanitize@url \@url }%
\providecommand \@url [1]{\endgroup\@href {#1}{\urlprefix }}%
\providecommand \urlprefix  [0]{URL }%
\providecommand \Eprint [0]{\href }%
\providecommand \doibase [0]{https://doi.org/}%
\providecommand \selectlanguage [0]{\@gobble}%
\providecommand \bibinfo  [0]{\@secondoftwo}%
\providecommand \bibfield  [0]{\@secondoftwo}%
\providecommand \translation [1]{[#1]}%
\providecommand \BibitemOpen [0]{}%
\providecommand \bibitemStop [0]{}%
\providecommand \bibitemNoStop [0]{.\EOS\space}%
\providecommand \EOS [0]{\spacefactor3000\relax}%
\providecommand \BibitemShut  [1]{\csname bibitem#1\endcsname}%
\let\auto@bib@innerbib\@empty
\bibitem [{\citenamefont {Ising}(1925)}]{RN75}%
  \BibitemOpen
  \bibfield  {author} {\bibinfo {author} {\bibfnamefont {E.}~\bibnamefont
  {Ising}},\ }\bibfield  {title} {\emph {\bibinfo {title} {{Beitrag zur Theorie
  des Ferromagnetismus}}},\ }\href {https://doi.org/10.1007/BF02980577}
  {\bibfield  {journal} {\bibinfo  {journal} {Z. Phys.}\ }\textbf {\bibinfo
  {volume} {31}},\ \bibinfo {pages} {253--258} (\bibinfo {year}
  {1925})}\BibitemShut {NoStop}%
\bibitem [{\citenamefont {Brush}(1967)}]{RN441}%
  \BibitemOpen
  \bibfield  {author} {\bibinfo {author} {\bibfnamefont {S.~G.}\ \bibnamefont
  {Brush}},\ }\bibfield  {title} {\emph {\bibinfo {title} {{History of the
  Lenz-Ising Model}}},\ }\href {https://doi.org/10.1103/RevModPhys.39.883}
  {\bibfield  {journal} {\bibinfo  {journal} {Rev. Mod. Phys.}\ }\textbf
  {\bibinfo {volume} {39}},\ \bibinfo {pages} {883--893} (\bibinfo {year}
  {1967})}\BibitemShut {NoStop}%
\bibitem [{\citenamefont {Niss}(2005)}]{RN312}%
  \BibitemOpen
  \bibfield  {author} {\bibinfo {author} {\bibfnamefont {M.}~\bibnamefont
  {Niss}},\ }\bibfield  {title} {\emph {\bibinfo {title} {{History of the
  Lenz-Ising Model 1920–1950: From Ferromagnetic to Cooperative
  Phenomena}}},\ }\href {https://doi.org/10.1007/s00407-004-0088-3} {\bibfield
  {journal} {\bibinfo  {journal} {Arch. Hist. Exact Sci.}\ }\textbf {\bibinfo
  {volume} {59}},\ \bibinfo {pages} {267--318} (\bibinfo {year}
  {2005})}\BibitemShut {NoStop}%
\bibitem [{\citenamefont {Peierls}(1936)}]{RN460}%
  \BibitemOpen
  \bibfield  {author} {\bibinfo {author} {\bibfnamefont {R.}~\bibnamefont
  {Peierls}},\ }\bibfield  {title} {\emph {\bibinfo {title} {{On Ising's model
  of ferromagnetism}}},\ }\href {https://doi.org/10.1017/S0305004100019174}
  {\bibfield  {journal} {\bibinfo  {journal} {Proc. Camb. Phil. Soc.}\ }\textbf
  {\bibinfo {volume} {32}},\ \bibinfo {pages} {477--481} (\bibinfo {year}
  {1936})}\BibitemShut {NoStop}%
\bibitem [{\citenamefont {Kramers}\ and\ \citenamefont
  {Wannier}(1941{\natexlab{a}})}]{RN236}%
  \BibitemOpen
  \bibfield  {author} {\bibinfo {author} {\bibfnamefont {H.~A.}\ \bibnamefont
  {Kramers}}\ and\ \bibinfo {author} {\bibfnamefont {G.~H.}\ \bibnamefont
  {Wannier}},\ }\bibfield  {title} {\emph {\bibinfo {title} {{Statistics of the
  Two-Dimensional Ferromagnet. Part I}}},\ }\href
  {https://doi.org/10.1103/PhysRev.60.252} {\bibfield  {journal} {\bibinfo
  {journal} {Phys. Rev.}\ }\textbf {\bibinfo {volume} {60}},\ \bibinfo {pages}
  {252--262} (\bibinfo {year} {1941}{\natexlab{a}})}\BibitemShut {NoStop}%
\bibitem [{\citenamefont {Kramers}\ and\ \citenamefont
  {Wannier}(1941{\natexlab{b}})}]{RN237}%
  \BibitemOpen
  \bibfield  {author} {\bibinfo {author} {\bibfnamefont {H.~A.}\ \bibnamefont
  {Kramers}}\ and\ \bibinfo {author} {\bibfnamefont {G.~H.}\ \bibnamefont
  {Wannier}},\ }\bibfield  {title} {\emph {\bibinfo {title} {{Statistics of the
  Two-Dimensional Ferromagnet. Part II}}},\ }\href
  {https://doi.org/10.1103/PhysRev.60.263} {\bibfield  {journal} {\bibinfo
  {journal} {Phys. Rev.}\ }\textbf {\bibinfo {volume} {60}},\ \bibinfo {pages}
  {263--276} (\bibinfo {year} {1941}{\natexlab{b}})}\BibitemShut {NoStop}%
\bibitem [{\citenamefont {Onsager}(1944)}]{RN72}%
  \BibitemOpen
  \bibfield  {author} {\bibinfo {author} {\bibfnamefont {L.}~\bibnamefont
  {Onsager}},\ }\bibfield  {title} {\emph {\bibinfo {title} {{Crystal
  Statistics. I. A Two-Dimensional Model with an Order-Disorder Transition}}},\
  }\href {https://doi.org/10.1103/PhysRev.65.117} {\bibfield  {journal}
  {\bibinfo  {journal} {Phys. Rev.}\ }\textbf {\bibinfo {volume} {65}},\
  \bibinfo {pages} {117--149} (\bibinfo {year} {1944})}\BibitemShut {NoStop}%
\bibitem [{\citenamefont {Husimi}\ and\ \citenamefont {Syôzi}(1950)}]{RN122}%
  \BibitemOpen
  \bibfield  {author} {\bibinfo {author} {\bibfnamefont {K.}~\bibnamefont
  {Husimi}}\ and\ \bibinfo {author} {\bibfnamefont {I.}~\bibnamefont
  {Syôzi}},\ }\bibfield  {title} {\emph {\bibinfo {title} {{The Statistics of
  Honeycomb and Triangular Lattice. I}}},\ }\href
  {https://doi.org/10.1143/ptp/5.2.177} {\bibfield  {journal} {\bibinfo
  {journal} {Prog. Theor. Phys.}\ }\textbf {\bibinfo {volume} {5}},\ \bibinfo
  {pages} {177--186} (\bibinfo {year} {1950})}\BibitemShut {NoStop}%
\bibitem [{\citenamefont {Syôzi}(1950)}]{RN123}%
  \BibitemOpen
  \bibfield  {author} {\bibinfo {author} {\bibfnamefont {I.}~\bibnamefont
  {Syôzi}},\ }\bibfield  {title} {\emph {\bibinfo {title} {{The Statistics of
  Honeycomb and Triangular Lattice. II}}},\ }\href
  {https://doi.org/10.1143/ptp/5.3.341} {\bibfield  {journal} {\bibinfo
  {journal} {Prog. Theor. Phys.}\ }\textbf {\bibinfo {volume} {5}},\ \bibinfo
  {pages} {341--351} (\bibinfo {year} {1950})}\BibitemShut {NoStop}%
\bibitem [{\citenamefont {Wannier}(1950)}]{RN81}%
  \BibitemOpen
  \bibfield  {author} {\bibinfo {author} {\bibfnamefont {G.~H.}\ \bibnamefont
  {Wannier}},\ }\bibfield  {title} {\emph {\bibinfo {title}
  {{Antiferromagnetism. The Triangular Ising Net}}},\ }\href
  {https://doi.org/10.1103/PhysRev.79.357} {\bibfield  {journal} {\bibinfo
  {journal} {Phys. Rev.}\ }\textbf {\bibinfo {volume} {79}},\ \bibinfo {pages}
  {357--364} (\bibinfo {year} {1950})}\BibitemShut {NoStop}%
\bibitem [{\citenamefont {Syôzi}(1951)}]{RN121}%
  \BibitemOpen
  \bibfield  {author} {\bibinfo {author} {\bibfnamefont {I.}~\bibnamefont
  {Syôzi}},\ }\bibfield  {title} {\emph {\bibinfo {title} {{Statistics of
  Kagomé Lattice}}},\ }\href {https://doi.org/10.1143/ptp/6.3.306} {\bibfield
  {journal} {\bibinfo  {journal} {Prog. Theor. Phys.}\ }\textbf {\bibinfo
  {volume} {6}},\ \bibinfo {pages} {306--308} (\bibinfo {year}
  {1951})}\BibitemShut {NoStop}%
\bibitem [{\citenamefont {Kanô}\ and\ \citenamefont {Naya}(1953)}]{RN82}%
  \BibitemOpen
  \bibfield  {author} {\bibinfo {author} {\bibfnamefont {K.}~\bibnamefont
  {Kanô}}\ and\ \bibinfo {author} {\bibfnamefont {S.}~\bibnamefont {Naya}},\
  }\bibfield  {title} {\emph {\bibinfo {title} {{Antiferromagnetism. The
  Kagomé Ising Net}}},\ }\href {https://doi.org/10.1143/ptp/10.2.158}
  {\bibfield  {journal} {\bibinfo  {journal} {Prog. Theor. Phys.}\ }\textbf
  {\bibinfo {volume} {10}},\ \bibinfo {pages} {158--172} (\bibinfo {year}
  {1953})}\BibitemShut {NoStop}%
\bibitem [{\citenamefont {Giacomini}(1985)}]{RN58}%
  \BibitemOpen
  \bibfield  {author} {\bibinfo {author} {\bibfnamefont {H.~J.}\ \bibnamefont
  {Giacomini}},\ }\bibfield  {title} {\emph {\bibinfo {title} {{Exact results
  for a checkerboard Ising model with crossing and four-spin interactions}}},\
  }\href {https://doi.org/10.1088/0305-4470/18/17/005} {\bibfield  {journal}
  {\bibinfo  {journal} {J. Phys. A: Math. Gen.}\ }\textbf {\bibinfo {volume}
  {18}},\ \bibinfo {pages} {L1087--L1093} (\bibinfo {year} {1985})}\BibitemShut
  {NoStop}%
\bibitem [{\citenamefont {Lee}\ and\ \citenamefont {Yang}(1952)}]{RN57}%
  \BibitemOpen
  \bibfield  {author} {\bibinfo {author} {\bibfnamefont {T.~D.}\ \bibnamefont
  {Lee}}\ and\ \bibinfo {author} {\bibfnamefont {C.~N.}\ \bibnamefont {Yang}},\
  }\bibfield  {title} {\emph {\bibinfo {title} {{Statistical Theory of
  Equations of State and Phase Transitions. II. Lattice Gas and Ising
  Model}}},\ }\href {https://doi.org/10.1103/PhysRev.87.410} {\bibfield
  {journal} {\bibinfo  {journal} {Phys. Rev.}\ }\textbf {\bibinfo {volume}
  {87}},\ \bibinfo {pages} {410--419} (\bibinfo {year} {1952})}\BibitemShut
  {NoStop}%
\bibitem [{\citenamefont {Baxter}(1965)}]{RN67}%
  \BibitemOpen
  \bibfield  {author} {\bibinfo {author} {\bibfnamefont {G.}~\bibnamefont
  {Baxter}},\ }\bibfield  {title} {\emph {\bibinfo {title} {{Weight Factors for
  the Two‐Dimensional Ising Model}}},\ }\href
  {https://doi.org/10.1063/1.1704362} {\bibfield  {journal} {\bibinfo
  {journal} {J. Math. Phys.}\ }\textbf {\bibinfo {volume} {6}},\ \bibinfo
  {pages} {1015--1021} (\bibinfo {year} {1965})}\BibitemShut {NoStop}%
\bibitem [{\citenamefont {McCoy}\ and\ \citenamefont
  {Wu}(1967{\natexlab{a}})}]{RN68}%
  \BibitemOpen
  \bibfield  {author} {\bibinfo {author} {\bibfnamefont {B.~M.}\ \bibnamefont
  {McCoy}}\ and\ \bibinfo {author} {\bibfnamefont {T.~T.}\ \bibnamefont {Wu}},\
  }\bibfield  {title} {\emph {\bibinfo {title} {{Theory of Toeplitz
  Determinants and the Spin Correlations of the Two-Dimensional Ising Model.
  II}}},\ }\href {https://doi.org/10.1103/PhysRev.155.438} {\bibfield
  {journal} {\bibinfo  {journal} {Phys. Rev.}\ }\textbf {\bibinfo {volume}
  {155}},\ \bibinfo {pages} {438--452} (\bibinfo {year}
  {1967}{\natexlab{a}})}\BibitemShut {NoStop}%
\bibitem [{\citenamefont {Wu}(1986)}]{RN51}%
  \BibitemOpen
  \bibfield  {author} {\bibinfo {author} {\bibfnamefont {F.~Y.}\ \bibnamefont
  {Wu}},\ }\bibfield  {title} {\emph {\bibinfo {title} {{Two-dimensional Ising
  model with crossing and four-spin interactions and a magnetic field
  i($\pi$/2)kT}}},\ }\href {https://doi.org/10.1007/BF01011305} {\bibfield
  {journal} {\bibinfo  {journal} {J. Stat. Phys.}\ }\textbf {\bibinfo {volume}
  {44}},\ \bibinfo {pages} {455--463} (\bibinfo {year} {1986})}\BibitemShut
  {NoStop}%
\bibitem [{\citenamefont {Lin}\ and\ \citenamefont {Wu}(1988)}]{RN274}%
  \BibitemOpen
  \bibfield  {author} {\bibinfo {author} {\bibfnamefont {K.~Y.}\ \bibnamefont
  {Lin}}\ and\ \bibinfo {author} {\bibfnamefont {F.~Y.}\ \bibnamefont {Wu}},\
  }\bibfield  {title} {\emph {\bibinfo {title} {{Ising Model In The Magnetic
  Field i$\pi$kT/2}}},\ }\href {https://doi.org/10.1142/S0217979288000330}
  {\bibfield  {journal} {\bibinfo  {journal} {Int. J. Mod. Phys. B}\ }\textbf
  {\bibinfo {volume} {02}},\ \bibinfo {pages} {471--481} (\bibinfo {year}
  {1988})}\BibitemShut {NoStop}%
\bibitem [{\citenamefont {Li}\ \emph {et~al.}(2025)\citenamefont {Li},
  \citenamefont {Wang},\ and\ \citenamefont {Yang}}]{RN558}%
  \BibitemOpen
  \bibfield  {author} {\bibinfo {author} {\bibfnamefont {D.-Z.}\ \bibnamefont
  {Li}}, \bibinfo {author} {\bibfnamefont {X.}~\bibnamefont {Wang}},\ and\
  \bibinfo {author} {\bibfnamefont {X.-B.}\ \bibnamefont {Yang}},\ }\bibfield
  {title} {\emph {\bibinfo {title} {{Free-Fermion Models and Two-Dimensional
  Ising Models Under Zero Field and Imaginary Field $i(\pi/2)k_BT$}}},\ }\href
  {https://doi.org/10.3390/e27080799} {\bibfield  {journal} {\bibinfo
  {journal} {Entropy}\ }\textbf {\bibinfo {volume} {27}},\ \bibinfo {pages}
  {799} (\bibinfo {year} {2025})}\BibitemShut {NoStop}%
\bibitem [{\citenamefont {Baxter}(1982)}]{RN49}%
  \BibitemOpen
  \bibfield  {author} {\bibinfo {author} {\bibfnamefont {R.~J.}\ \bibnamefont
  {Baxter}},\ }\href@noop {} {\emph {\bibinfo {title} {{Exactly solved models
  in statistical mechanics}}}}\ (\bibinfo  {publisher} {Academic Press},\
  \bibinfo {address} {London},\ \bibinfo {year} {1982})\BibitemShut {NoStop}%
\bibitem [{\citenamefont {McCoy}\ and\ \citenamefont {Wu}(2014)}]{RN465}%
  \BibitemOpen
  \bibfield  {author} {\bibinfo {author} {\bibfnamefont {B.}~\bibnamefont
  {McCoy}}\ and\ \bibinfo {author} {\bibfnamefont {T.}~\bibnamefont {Wu}},\
  }\href@noop {} {\emph {\bibinfo {title} {{The Two-Dimensional Ising Model:
  Second Edition}}}}\ (\bibinfo  {publisher} {Dover Publications},\ \bibinfo
  {address} {New York},\ \bibinfo {year} {2014})\BibitemShut {NoStop}%
\bibitem [{\citenamefont {Kaufman}(1949)}]{RN73}%
  \BibitemOpen
  \bibfield  {author} {\bibinfo {author} {\bibfnamefont {B.}~\bibnamefont
  {Kaufman}},\ }\bibfield  {title} {\emph {\bibinfo {title} {{Crystal
  Statistics. II. Partition Function Evaluated by Spinor Analysis}}},\ }\href
  {https://doi.org/10.1103/PhysRev.76.1232} {\bibfield  {journal} {\bibinfo
  {journal} {Phys. Rev.}\ }\textbf {\bibinfo {volume} {76}},\ \bibinfo {pages}
  {1232--1243} (\bibinfo {year} {1949})}\BibitemShut {NoStop}%
\bibitem [{\citenamefont {Nambu}(1950)}]{RN267}%
  \BibitemOpen
  \bibfield  {author} {\bibinfo {author} {\bibfnamefont {Y.}~\bibnamefont
  {Nambu}},\ }\bibfield  {title} {\emph {\bibinfo {title} {{A Note on the
  Eigenvalue Problem in Crystal Statistics}}},\ }\href
  {https://doi.org/10.1143/ptp/5.1.1} {\bibfield  {journal} {\bibinfo
  {journal} {Prog. Theor. Phys.}\ }\textbf {\bibinfo {volume} {5}},\ \bibinfo
  {pages} {1--13} (\bibinfo {year} {1950})}\BibitemShut {NoStop}%
\bibitem [{\citenamefont {Schultz}\ \emph {et~al.}(1964)\citenamefont
  {Schultz}, \citenamefont {Mattis},\ and\ \citenamefont {Lieb}}]{RN76}%
  \BibitemOpen
  \bibfield  {author} {\bibinfo {author} {\bibfnamefont {T.~D.}\ \bibnamefont
  {Schultz}}, \bibinfo {author} {\bibfnamefont {D.~C.}\ \bibnamefont
  {Mattis}},\ and\ \bibinfo {author} {\bibfnamefont {E.~H.}\ \bibnamefont
  {Lieb}},\ }\bibfield  {title} {\emph {\bibinfo {title} {{Two-Dimensional
  Ising Model as a Soluble Problem of Many Fermions}}},\ }\href
  {https://doi.org/10.1103/RevModPhys.36.856} {\bibfield  {journal} {\bibinfo
  {journal} {Rev. Mod. Phys.}\ }\textbf {\bibinfo {volume} {36}},\ \bibinfo
  {pages} {856--871} (\bibinfo {year} {1964})}\BibitemShut {NoStop}%
\bibitem [{\citenamefont {Thompson}(1965)}]{RN668}%
  \BibitemOpen
  \bibfield  {author} {\bibinfo {author} {\bibfnamefont {C.~J.}\ \bibnamefont
  {Thompson}},\ }\bibfield  {title} {\emph {\bibinfo {title} {{Algebraic
  Derivation of the Partition Function of a Two‐Dimensional Ising Model}}},\
  }\href {https://doi.org/10.1063/1.1704789} {\bibfield  {journal} {\bibinfo
  {journal} {J. Math. Phys.}\ }\textbf {\bibinfo {volume} {6}},\ \bibinfo
  {pages} {1392--1395} (\bibinfo {year} {1965})}\BibitemShut {NoStop}%
\bibitem [{\citenamefont {Kastening}(2001)}]{RN666}%
  \BibitemOpen
  \bibfield  {author} {\bibinfo {author} {\bibfnamefont {B.}~\bibnamefont
  {Kastening}},\ }\bibfield  {title} {\emph {\bibinfo {title} {{Simplifying
  Kaufman's solution of the two-dimensional Ising model}}},\ }\href
  {https://doi.org/10.1103/PhysRevE.64.066106} {\bibfield  {journal} {\bibinfo
  {journal} {Phys. Rev. E}\ }\textbf {\bibinfo {volume} {64}},\ \bibinfo
  {pages} {066106} (\bibinfo {year} {2001})}\BibitemShut {NoStop}%
\bibitem [{\citenamefont {Kac}\ and\ \citenamefont {Ward}(1952)}]{RN74}%
  \BibitemOpen
  \bibfield  {author} {\bibinfo {author} {\bibfnamefont {M.}~\bibnamefont
  {Kac}}\ and\ \bibinfo {author} {\bibfnamefont {J.~C.}\ \bibnamefont {Ward}},\
  }\bibfield  {title} {\emph {\bibinfo {title} {{A Combinatorial Solution of
  the Two-Dimensional Ising Model}}},\ }\href
  {https://doi.org/10.1103/PhysRev.88.1332} {\bibfield  {journal} {\bibinfo
  {journal} {Phys. Rev.}\ }\textbf {\bibinfo {volume} {88}},\ \bibinfo {pages}
  {1332--1337} (\bibinfo {year} {1952})}\BibitemShut {NoStop}%
\bibitem [{\citenamefont {Potts}\ and\ \citenamefont {Ward}(1955)}]{RN272}%
  \BibitemOpen
  \bibfield  {author} {\bibinfo {author} {\bibfnamefont {R.~B.}\ \bibnamefont
  {Potts}}\ and\ \bibinfo {author} {\bibfnamefont {J.~C.}\ \bibnamefont
  {Ward}},\ }\bibfield  {title} {\emph {\bibinfo {title} {{The Combinatrial
  Method and the Two-Dimensional Ising Model}}},\ }\href
  {https://doi.org/10.1143/PTP.13.38} {\bibfield  {journal} {\bibinfo
  {journal} {Prog. Theor. Phys.}\ }\textbf {\bibinfo {volume} {13}},\ \bibinfo
  {pages} {38--46} (\bibinfo {year} {1955})}\BibitemShut {NoStop}%
\bibitem [{\citenamefont {Sherman}(1960)}]{RN611}%
  \BibitemOpen
  \bibfield  {author} {\bibinfo {author} {\bibfnamefont {S.}~\bibnamefont
  {Sherman}},\ }\bibfield  {title} {\emph {\bibinfo {title} {{Combinatorial
  Aspects of the Ising Model for Ferromagnetism. I. A Conjecture of Feynman on
  Paths and Graphs}}},\ }\href {https://doi.org/10.1063/1.1703653} {\bibfield
  {journal} {\bibinfo  {journal} {J. Math. Phys.}\ }\textbf {\bibinfo {volume}
  {1}},\ \bibinfo {pages} {202--217} (\bibinfo {year} {1960})}\BibitemShut
  {NoStop}%
\bibitem [{\citenamefont {Sherman}(1962)}]{RN662}%
  \BibitemOpen
  \bibfield  {author} {\bibinfo {author} {\bibfnamefont {S.}~\bibnamefont
  {Sherman}},\ }\bibfield  {title} {\emph {\bibinfo {title} {{Combinatorial
  aspects of the Ising model for ferromagnetism. II. An analogue to the Witt
  identity}}},\ }\href {https://doi.org/10.1090/S0002-9904-1962-10756-3}
  {\bibfield  {journal} {\bibinfo  {journal} {Bull. Amer. Math. Soc.}\ }\textbf
  {\bibinfo {volume} {68}},\ \bibinfo {pages} {225--229} (\bibinfo {year}
  {1962})}\BibitemShut {NoStop}%
\bibitem [{\citenamefont {Burgoyne}(1963)}]{RN663}%
  \BibitemOpen
  \bibfield  {author} {\bibinfo {author} {\bibfnamefont {P.~N.}\ \bibnamefont
  {Burgoyne}},\ }\bibfield  {title} {\emph {\bibinfo {title} {{Remarks on the
  Combinatorial Approach to the Ising Problem}}},\ }\href
  {https://doi.org/10.1063/1.1703907} {\bibfield  {journal} {\bibinfo
  {journal} {J. Math. Phys.}\ }\textbf {\bibinfo {volume} {4}},\ \bibinfo
  {pages} {1320--1326} (\bibinfo {year} {1963})}\BibitemShut {NoStop}%
\bibitem [{\citenamefont {Vdovichenko}(1965)}]{RN672}%
  \BibitemOpen
  \bibfield  {author} {\bibinfo {author} {\bibfnamefont {N.~V.}\ \bibnamefont
  {Vdovichenko}},\ }\bibfield  {title} {\emph {\bibinfo {title} {{A calculation
  of the partition function for a plane dipole lattice}}},\ }\href
  {http://jetp.ras.ru/cgi-bin/e/index/e/20/2/p477?a=list} {\bibfield  {journal}
  {\bibinfo  {journal} {Sov. Phys. JETP}\ }\textbf {\bibinfo {volume} {20}},\
  \bibinfo {pages} {477--479} (\bibinfo {year} {1965})}\BibitemShut {NoStop}%
\bibitem [{\citenamefont {Glasser}(1970)}]{RN664}%
  \BibitemOpen
  \bibfield  {author} {\bibinfo {author} {\bibfnamefont {M.~L.}\ \bibnamefont
  {Glasser}},\ }\bibfield  {title} {\emph {\bibinfo {title} {{Exact Partition
  Function for the Two-Dimensional Ising Model}}},\ }\href
  {https://doi.org/10.1119/1.1976530} {\bibfield  {journal} {\bibinfo
  {journal} {Am. J. Phys.}\ }\textbf {\bibinfo {volume} {38}},\ \bibinfo
  {pages} {1033--1036} (\bibinfo {year} {1970})}\BibitemShut {NoStop}%
\bibitem [{\citenamefont {da~Costa}\ and\ \citenamefont
  {Maciel}(2003)}]{RN673}%
  \BibitemOpen
  \bibfield  {author} {\bibinfo {author} {\bibfnamefont {G.}~\bibnamefont
  {da~Costa}}\ and\ \bibinfo {author} {\bibfnamefont {A.~L.}\ \bibnamefont
  {Maciel}},\ }\bibfield  {title} {\emph {\bibinfo {title} {{Combinatorial
  formulation of Ising model revisited}}},\ }\href
  {https://doi.org/10.1590/S1806-11172003000100007} {\bibfield  {journal}
  {\bibinfo  {journal} {Rev. Bras. Ensino Fís.}\ }\textbf {\bibinfo {volume}
  {25}},\ \bibinfo {pages} {49--61} (\bibinfo {year} {2003})}\BibitemShut
  {NoStop}%
\bibitem [{\citenamefont {Hurst}\ and\ \citenamefont {Green}(1960)}]{RN207}%
  \BibitemOpen
  \bibfield  {author} {\bibinfo {author} {\bibfnamefont {C.~A.}\ \bibnamefont
  {Hurst}}\ and\ \bibinfo {author} {\bibfnamefont {H.~S.}\ \bibnamefont
  {Green}},\ }\bibfield  {title} {\emph {\bibinfo {title} {{New Solution of the
  Ising Problem for a Rectangular Lattice}}},\ }\href
  {https://doi.org/10.1063/1.1731333} {\bibfield  {journal} {\bibinfo
  {journal} {J. Chem. Phys.}\ }\textbf {\bibinfo {volume} {33}},\ \bibinfo
  {pages} {1059--1062} (\bibinfo {year} {1960})}\BibitemShut {NoStop}%
\bibitem [{\citenamefont {Hurst}(1964)}]{RN480}%
  \BibitemOpen
  \bibfield  {author} {\bibinfo {author} {\bibfnamefont {C.~A.}\ \bibnamefont
  {Hurst}},\ }\bibfield  {title} {\emph {\bibinfo {title} {{Applicability of
  the Pfaffian Method to Combinatorial Problems on a Lattice}}},\ }\href
  {https://doi.org/10.1063/1.1704068} {\bibfield  {journal} {\bibinfo
  {journal} {J. Math. Phys.}\ }\textbf {\bibinfo {volume} {5}},\ \bibinfo
  {pages} {90--100} (\bibinfo {year} {1964})}\BibitemShut {NoStop}%
\bibitem [{\citenamefont {Hurst}(1966)}]{RN265}%
  \BibitemOpen
  \bibfield  {author} {\bibinfo {author} {\bibfnamefont {C.~A.}\ \bibnamefont
  {Hurst}},\ }\bibfield  {title} {\emph {\bibinfo {title} {{New Approach to the
  Ising Problem}}},\ }\href {https://doi.org/10.1063/1.1704933} {\bibfield
  {journal} {\bibinfo  {journal} {J. Math. Phys.}\ }\textbf {\bibinfo {volume}
  {7}},\ \bibinfo {pages} {305--310} (\bibinfo {year} {1966})}\BibitemShut
  {NoStop}%
\bibitem [{\citenamefont {Fisher}(1966)}]{RN218}%
  \BibitemOpen
  \bibfield  {author} {\bibinfo {author} {\bibfnamefont {M.~E.}\ \bibnamefont
  {Fisher}},\ }\bibfield  {title} {\emph {\bibinfo {title} {{On the Dimer
  Solution of Planar Ising Models}}},\ }\href
  {https://doi.org/10.1063/1.1704825} {\bibfield  {journal} {\bibinfo
  {journal} {J. Math. Phys.}\ }\textbf {\bibinfo {volume} {7}},\ \bibinfo
  {pages} {1776--1781} (\bibinfo {year} {1966})}\BibitemShut {NoStop}%
\bibitem [{\citenamefont {Gibberd}\ and\ \citenamefont {Hurst}(1967)}]{RN269}%
  \BibitemOpen
  \bibfield  {author} {\bibinfo {author} {\bibfnamefont {R.~W.}\ \bibnamefont
  {Gibberd}}\ and\ \bibinfo {author} {\bibfnamefont {C.~A.}\ \bibnamefont
  {Hurst}},\ }\bibfield  {title} {\emph {\bibinfo {title} {{New Approach to the
  Ising ModeI. II}}},\ }\href {https://doi.org/10.1063/1.1705355} {\bibfield
  {journal} {\bibinfo  {journal} {J. Math. Phys.}\ }\textbf {\bibinfo {volume}
  {8}},\ \bibinfo {pages} {1427--1435} (\bibinfo {year} {1967})}\BibitemShut
  {NoStop}%
\bibitem [{\citenamefont {Montroll}(1964)}]{RN397}%
  \BibitemOpen
  \bibfield  {author} {\bibinfo {author} {\bibfnamefont {E.~W.}\ \bibnamefont
  {Montroll}},\ }\bibinfo {title} {{Lattice Statistics}},\ in\ \href@noop {}
  {\emph {\bibinfo {booktitle} {Applied Combinatorial Mathematics}}}\ (\bibinfo
   {publisher} {Wiley},\ \bibinfo {address} {New York},\ \bibinfo {year}
  {1964})\BibitemShut {NoStop}%
\bibitem [{\citenamefont {Samuel}(1980)}]{RN209}%
  \BibitemOpen
  \bibfield  {author} {\bibinfo {author} {\bibfnamefont {S.}~\bibnamefont
  {Samuel}},\ }\bibfield  {title} {\emph {\bibinfo {title} {{The use of
  anticommuting variable integrals in statistical mechanics. I. The computation
  of partition functions}}},\ }\href {https://doi.org/10.1063/1.524404}
  {\bibfield  {journal} {\bibinfo  {journal} {J. Math. Phys.}\ }\textbf
  {\bibinfo {volume} {21}},\ \bibinfo {pages} {2806--2814} (\bibinfo {year}
  {1980})}\BibitemShut {NoStop}%
\bibitem [{\citenamefont {Plechko}(1985)}]{RN674}%
  \BibitemOpen
  \bibfield  {author} {\bibinfo {author} {\bibfnamefont {V.~N.}\ \bibnamefont
  {Plechko}},\ }\bibfield  {title} {\emph {\bibinfo {title} {{Simple solution
  of two-dimensional ising model on a torus in terms of Grassmann
  integrals}}},\ }\href {https://doi.org/10.1007/BF01017042} {\bibfield
  {journal} {\bibinfo  {journal} {Theor. Math. Phys.}\ }\textbf {\bibinfo
  {volume} {64}},\ \bibinfo {pages} {748--756} (\bibinfo {year}
  {1985})}\BibitemShut {NoStop}%
\bibitem [{\citenamefont {Brascamp}\ and\ \citenamefont {Kunz}(1974)}]{RN410}%
  \BibitemOpen
  \bibfield  {author} {\bibinfo {author} {\bibfnamefont {H.~J.}\ \bibnamefont
  {Brascamp}}\ and\ \bibinfo {author} {\bibfnamefont {H.}~\bibnamefont
  {Kunz}},\ }\bibfield  {title} {\emph {\bibinfo {title} {{Zeroes of the
  partition function for the Ising model in the complex temperature plane}}},\
  }\href {https://doi.org/10.1063/1.1666505} {\bibfield  {journal} {\bibinfo
  {journal} {J. Math. Phys.}\ }\textbf {\bibinfo {volume} {15}},\ \bibinfo
  {pages} {65--66} (\bibinfo {year} {1974})}\BibitemShut {NoStop}%
\bibitem [{\citenamefont {Fisher}(1965)}]{RN308}%
  \BibitemOpen
  \bibfield  {author} {\bibinfo {author} {\bibfnamefont {M.~E.}\ \bibnamefont
  {Fisher}},\ }\bibinfo {title} {{The nature of critical points}},\ in\
  \href@noop {} {\emph {\bibinfo {booktitle} {Lectures in Theoretical Physics:
  Volume VII C - Statistical Physics, Weak Interactions, Field Theory}}}\
  (\bibinfo  {publisher} {University of Colorado Press},\ \bibinfo {address}
  {Boulder},\ \bibinfo {year} {1965})\BibitemShut {NoStop}%
\bibitem [{\citenamefont {Lu}\ and\ \citenamefont {Wu}(2001)}]{RN411}%
  \BibitemOpen
  \bibfield  {author} {\bibinfo {author} {\bibfnamefont {W.~T.}\ \bibnamefont
  {Lu}}\ and\ \bibinfo {author} {\bibfnamefont {F.~Y.}\ \bibnamefont {Wu}},\
  }\bibfield  {title} {\emph {\bibinfo {title} {{Density of the Fisher Zeroes
  for the Ising Model}}},\ }\href {https://doi.org/10.1023/A:1004863322373}
  {\bibfield  {journal} {\bibinfo  {journal} {J. Stat. Phys.}\ }\textbf
  {\bibinfo {volume} {102}},\ \bibinfo {pages} {953--970} (\bibinfo {year}
  {2001})}\BibitemShut {NoStop}%
\bibitem [{\citenamefont {McCoy}\ and\ \citenamefont
  {Wu}(1967{\natexlab{b}})}]{RN459}%
  \BibitemOpen
  \bibfield  {author} {\bibinfo {author} {\bibfnamefont {B.~M.}\ \bibnamefont
  {McCoy}}\ and\ \bibinfo {author} {\bibfnamefont {T.~T.}\ \bibnamefont {Wu}},\
  }\bibfield  {title} {\emph {\bibinfo {title} {{Theory of Toeplitz
  Determinants and the Spin Correlations of the Two-Dimensional Ising Model.
  IV}}},\ }\href {https://doi.org/10.1103/PhysRev.162.436} {\bibfield
  {journal} {\bibinfo  {journal} {Phys. Rev.}\ }\textbf {\bibinfo {volume}
  {162}},\ \bibinfo {pages} {436--475} (\bibinfo {year}
  {1967}{\natexlab{b}})}\BibitemShut {NoStop}%
\bibitem [{\citenamefont {Kastening}(2002)}]{RN474}%
  \BibitemOpen
  \bibfield  {author} {\bibinfo {author} {\bibfnamefont {B.}~\bibnamefont
  {Kastening}},\ }\bibfield  {title} {\emph {\bibinfo {title} {{Simplified
  transfer matrix approach in the two-dimensional Ising model with various
  boundary conditions}}},\ }\href {https://doi.org/10.1103/PhysRevE.66.057103}
  {\bibfield  {journal} {\bibinfo  {journal} {Phys. Rev. E}\ }\textbf {\bibinfo
  {volume} {66}},\ \bibinfo {pages} {057103} (\bibinfo {year}
  {2002})}\BibitemShut {NoStop}%
\bibitem [{\citenamefont {Lyberg}(2008)}]{RN432}%
  \BibitemOpen
  \bibfield  {author} {\bibinfo {author} {\bibfnamefont {I.}~\bibnamefont
  {Lyberg}},\ }\bibfield  {title} {\emph {\bibinfo {title} {{The Ising lattice
  with Brascamp-Kunz boundary conditions and an external magnetic field}}},\
  }\bibfield  {journal} {\bibinfo  {journal} {arXiv preprint arXiv:0805.2497}\
  }\href {https://doi.org/10.48550/arXiv.0805.2497} {10.48550/arXiv.0805.2497}
  (\bibinfo {year} {2008})\BibitemShut {NoStop}%
\bibitem [{\citenamefont {Lyberg}(2013)}]{RN455}%
  \BibitemOpen
  \bibfield  {author} {\bibinfo {author} {\bibfnamefont {I.}~\bibnamefont
  {Lyberg}},\ }\bibfield  {title} {\emph {\bibinfo {title} {{Free energy of the
  anisotropic Ising lattice with Brascamp-Kunz boundary conditions}}},\ }\href
  {https://doi.org/10.1103/PhysRevE.87.062141} {\bibfield  {journal} {\bibinfo
  {journal} {Phys. Rev. E}\ }\textbf {\bibinfo {volume} {87}},\ \bibinfo
  {pages} {062141} (\bibinfo {year} {2013})}\BibitemShut {NoStop}%
\bibitem [{\citenamefont {Li}\ and\ \citenamefont {Wang}(2025)}]{RN671}%
  \BibitemOpen
  \bibfield  {author} {\bibinfo {author} {\bibfnamefont {D.-Z.}\ \bibnamefont
  {Li}}\ and\ \bibinfo {author} {\bibfnamefont {X.}~\bibnamefont {Wang}},\
  }\bibfield  {title} {\emph {\bibinfo {title} {{Free-fermion approach to the
  partition function zeros: Special boundary conditions and product form of
  solution}}},\ }\href {https://doi.org/10.1103/b6d1-6sk5} {\bibfield
  {journal} {\bibinfo  {journal} {Phys. Rev. Research}\ }\textbf {\bibinfo
  {volume} {7}},\ \bibinfo {pages} {043258} (\bibinfo {year}
  {2025})}\BibitemShut {NoStop}%
\bibitem [{\citenamefont {Abraham}(1971)}]{RN638}%
  \BibitemOpen
  \bibfield  {author} {\bibinfo {author} {\bibfnamefont {D.~B.}\ \bibnamefont
  {Abraham}},\ }\bibfield  {title} {\emph {\bibinfo {title} {{On the Transfer
  Matrix for the Two-Dimensional Ising Model}}},\ }\href
  {https://doi.org/10.1002/sapm197150171} {\bibfield  {journal} {\bibinfo
  {journal} {Stud. Appl. Math.}\ }\textbf {\bibinfo {volume} {50}},\ \bibinfo
  {pages} {71--88} (\bibinfo {year} {1971})}\BibitemShut {NoStop}%
\bibitem [{\citenamefont {Izmailian}\ and\ \citenamefont {Yeh}(2009)}]{RN659}%
  \BibitemOpen
  \bibfield  {author} {\bibinfo {author} {\bibfnamefont {N.~S.}\ \bibnamefont
  {Izmailian}}\ and\ \bibinfo {author} {\bibfnamefont {Y.-N.}\ \bibnamefont
  {Yeh}},\ }\bibfield  {title} {\emph {\bibinfo {title} {{Ising model with
  mixed boundary conditions: Universal amplitude ratios}}},\ }\href
  {https://doi.org/10.1016/j.nuclphysb.2009.01.017} {\bibfield  {journal}
  {\bibinfo  {journal} {Nucl. Phys. B}\ }\textbf {\bibinfo {volume} {814}},\
  \bibinfo {pages} {573--581} (\bibinfo {year} {2009})}\BibitemShut {NoStop}%
\bibitem [{\citenamefont {Baxter}(2017)}]{RN508}%
  \BibitemOpen
  \bibfield  {author} {\bibinfo {author} {\bibfnamefont {R.~J.}\ \bibnamefont
  {Baxter}},\ }\bibfield  {title} {\emph {\bibinfo {title} {{The bulk, surface
  and corner free energies of the square lattice Ising model}}},\ }\href
  {https://doi.org/10.1088/1751-8113/50/1/014001} {\bibfield  {journal}
  {\bibinfo  {journal} {J. Phys. A: Math. Theor.}\ }\textbf {\bibinfo {volume}
  {50}},\ \bibinfo {pages} {014001} (\bibinfo {year} {2017})}\BibitemShut
  {NoStop}%
\bibitem [{\citenamefont {Poghosyan}\ \emph {et~al.}(2017)\citenamefont
  {Poghosyan}, \citenamefont {Izmailian},\ and\ \citenamefont {Kenna}}]{RN637}%
  \BibitemOpen
  \bibfield  {author} {\bibinfo {author} {\bibfnamefont {A.}~\bibnamefont
  {Poghosyan}}, \bibinfo {author} {\bibfnamefont {N.}~\bibnamefont
  {Izmailian}},\ and\ \bibinfo {author} {\bibfnamefont {R.}~\bibnamefont
  {Kenna}},\ }\bibfield  {title} {\emph {\bibinfo {title} {{Exact solution of
  the critical Ising model with special toroidal boundary conditions}}},\
  }\href {https://doi.org/10.1103/PhysRevE.96.062127} {\bibfield  {journal}
  {\bibinfo  {journal} {Phys. Rev. E}\ }\textbf {\bibinfo {volume} {96}},\
  \bibinfo {pages} {062127} (\bibinfo {year} {2017})}\BibitemShut {NoStop}%
\bibitem [{\citenamefont {Jordan}\ and\ \citenamefont {Wigner}(1928)}]{RN497}%
  \BibitemOpen
  \bibfield  {author} {\bibinfo {author} {\bibfnamefont {P.}~\bibnamefont
  {Jordan}}\ and\ \bibinfo {author} {\bibfnamefont {E.}~\bibnamefont
  {Wigner}},\ }\bibfield  {title} {\emph {\bibinfo {title} {{Über das
  Paulische Äquivalenzverbot}}},\ }\href {https://doi.org/10.1007/BF01331938}
  {\bibfield  {journal} {\bibinfo  {journal} {Z. Phys.}\ }\textbf {\bibinfo
  {volume} {47}},\ \bibinfo {pages} {631--651} (\bibinfo {year}
  {1928})}\BibitemShut {NoStop}%
\bibitem [{\citenamefont {Beichert}(2013)}]{RN437}%
  \BibitemOpen
  \bibfield  {author} {\bibinfo {author} {\bibfnamefont {F.}~\bibnamefont
  {Beichert}},\ }\emph {\bibinfo {title} {{Phases at complex temperature :
  spiral correlation functions and regions of Fisher zeros for Ising
  models}}},\ \href@noop {} {\bibinfo {type} {Ph.{D}. thesis}} (\bibinfo {year}
  {University of St Andrews, 2013})\BibitemShut {NoStop}%
\end{thebibliography}
\end{document}